\documentclass[12pt]{article}

\usepackage{newtxtext,newtxmath}

\usepackage{graphicx}

\usepackage[letterpaper,margin=1in]{geometry}

\renewenvironment{abstract}
	{\quotation}
	{\endquotation}

\date{}

\makeatletter
\renewcommand{\fnum@figure}{\textbf{Figure \thefigure}}
\renewcommand{\fnum@table}{\textbf{Table \thetable}}
\makeatother

\usepackage{scicite}

\usepackage{url}

\usepackage{overpic}

\def\scititle{
	AT 2024wpp: the most luminous fast-evolving optical transient linked to the merger explosion of a black-hole binary
}
\title{\bfseries \boldmath \scititle}

\author
{Jialian Liu$^{1\dagger}$, 
Bao Wang$^{2,3\dagger}$, 
Xiaofeng Wang$^{1\dagger\ast}$, 
David Aguado$^{4,5}$,
Weili Lin$^{6}$,\and
Nancy-Elias Rosa$^{8,9}$, 
Qichun Liu$^{1}$, 
Frederick Poidevin$^{4,5}$,
Ismael Perez-Fournon$^{4,5}$,\and 
Long Li$^{7}$, 
Ailing Wang$^{10}$,
Yi Yang$^{1}$, 
Zigao Dai$^{2\ast}$, 
Alexei V. Filippenko$^{11,12}$,\and 
Thomas G. Brink$^{11}$, 
Di Xiao$^{3}$, 
Wenxiong Li$^{13}$, 
Yifang Liang$^{3}$,
Xuefeng Wu$^{3,2}$,\and 
Samaporn Tinyanont$^{14}$,
Jinjun Geng$^{3}$, 
Shengyu Yan$^{1}$, 
Weimin Yuan$^{13}$,
Jujia Zhang$^{15,16,17}$,\and 
Xiangyun Zeng$^{18,19}$, 
WeiKang Zheng$^{11}$, 
Yuanming Wang$^{20}$,
Tao An$^{21}$, 
YongZhi Cai$^{15,16,17}$,\and
Jeff Cooke$^{20,22}$, 
Lixin Dai$^{23,24}$, 
Andrea Farina$^{25}$, 
Maokai Hu$^{1}$,
Ye Li$^{3}$,
Chichuan Jin$^{13}$,\and 
Yuan Liu$^{13}$, 
David Lopez Fernandez-Nespral$^{4,5}$, 
Alicia Lopez Oramas$^{4,5}$,\and 
Andrea Reguitti$^{8,26}$, 
Xinwen Shu$^{27}$, 
Cuiying Song$^{1}$, 
Hui Sun$^{13}$,\and 
Ning-chen Sun$^{28,13,29}$, 
Lifan Wang$^{30}$, 
Tinggui Wang$^{2}$, 
Junjie Wei$^{3}$,
Qingyu Wu$^{13}$,\and
Danfeng Xiang$^{31}$,
Lei Yang$^{27}$,
Liping Li$^{15,16,17}$,
and Zhenyu Wang$^{15,16,17,28}$\and
\small$^{1}$Physics Department, Tsinghua University, Qinghua Yuan, Beijing, 100084, China \and
\small$^{2}$School of Astronomy and Space Sciences, University of Science and Technology of China, Hefei, 230026, China\and
\small$^{3}$Purple Mountain Observatory, Chinese Academy of Sciences, Nanjing, 210023, China\and
\small$^{4}$Instituto de Astrofisica de Canarias, Via Lactea, 38205 La Laguna, Tenerife, Spain\and
\small$^{5}$Universidad de La Laguna, Departamento de Astrof\'isica, 38206 La Laguna, Tenerife, Spain\and
\small$^{6}$Department of Astronomy, Xiamen University, Xiamen 361005, China\and
\small$^{7}$Department of Physics, School of Physics and Materials Science, Nanchang University, Nanchang, 330031, China\and
\small$^{8}$INAF -- Osservatorio Astronomico di Padova, Vicolo dell’Osservatorio 5, 35122 Padova, Italy\and
\small$^{9}$Institute of Space Sciences (ICE, CSIC), Campus UAB, Carrer de Can Magrans s/n, 08193 Barcelona, Spain\and
\small$^{10}$Institute of High Energy Physics, Chinese Academy of Sciences, Beijing, 100049, China\and
\small$^{11}$Department of Astronomy, University of California, Berkeley, CA 94720-3411, USA\and
\small$^{12}$Hagler Institute for Advanced Study, Texas A\&M University, 3572 TAMU, College Station, TX 77843, USA\and
\small$^{13}$National Astronomical Observatories, Chinese Academy of Sciences, 100101, China\and
\small$^{14}$National Astronomical Research Institute of Thailand, 260 Moo 4, Donkaew, Maerim, Chiang Mai, 50180, Thailand\and
\small$^{15}$Yunnan Observatories, Chinese Academy of Sciences, Kunming, 650216, China \and
\small$^{16}$International Centre of Supernovae, Yunnan Key Laboratory, Kunming, 650216, China \and
\small$^{17}$Key Laboratory for the Structure and Evolution of Celestial Objects, Chinese Academy of Sciences, \\ \small Kunming, 650216, China\and
\small$^{18}$Center for Astronomy and Space Sciences, China Three Gorges University, YiChang, 443000, China \and
\small$^{19}$College of Mathematics and Physics, China Three Gorges University, YiChang, 443000, China \and
\small$^{20}$Centre for Astrophysics and Supercomputing, Swinburne University of Technology, Hawthorn, VIC 3122, Australia\and
\small$^{21}$Shanghai Astronomical Observatory Chinese Academy of Sciences, Shanghai, 200030, China \and
\small$^{22}$ARC Centre of Excellence for Gravitational Wave Discovery (OzGrav), Swinburne University of Technology, \\ \small Hawthorn, VIC, 3122, Australia\and
\small$^{23}$Department of Physics, The University of Hong Kong, Pokfulam Road, Hong Kong, China\and
\small$^{24}$The Hong Kong Institute for Astronomy and Astrophysics, The University of Hong Kong, Pokfulam Road, \\ \small Hong Kong, China\and
\small$^{25}$Universit\`a degli Studi di Padova, Dipartimento di Fisica e Astronomia, Vicolo dell'Osservatorio 3, 35122 Padova, Italy\and 
\small$^{26}$INAF - Osservatorio Astronomico di Brera, Via Bianchi 46, 23807 Merate (LC), Italy\and
\small$^{27}$School of Physics and Electronic Information, Anhui Normal University, WuHu, 241008, China\and
\small$^{28}$School of Astronomy and Space Science, University of Chinese Academy of Sciences, Beijing, 100049, China\and
\small$^{29}$Institute for Frontiers in Astronomy and Astrophysics, Beijing Normal University, Beijing, 102206, China\and
\small$^{30}$George P. and Cynthia Woods Mitchell Institute for Fundamental Physics \& Astronomy, Texas A. \& M. University, \\ \small 4242 TAMU, College Station, TX 77843, USA\and
\small$^{31}$Beijing Planetarium, Beijing Academy of Science and Technology, Beijing 100044, China\and
\small$^\dagger$These authors contributed equally to this work\and
\small$^\ast$Corresponding authors:  wang\_xf@mail.tsinghua.edu.cn; daizg@ustc.edu.cn
}

\begin{document} 

\maketitle

\begin{abstract} \bfseries \boldmath
Fast blue optical transients (FBOTs) represent one of the most exotic astrophysical transients, exhibiting unusually strong emission across X-ray, optical, and radio wavelengths. Their physical origins remain highly debated, with proposed explanations ranging from stellar explosion to tidal disruption event (TDE). Here we report observations of the most luminous FBOT, AT 2024wpp whose post-peak luminosity rebrightens in X ray and becomes flattening in optical in a manner follows the decay rate characteristic of TDEs ($L_{\rm bol} \propto t^{-5/3}$). This invokes energy contribution of accretion by a central compact object, getting further corroborations from hardening of X-ray spectral index and detection of outflow inferred from the emission lines at similar phase. Detailed modeling of luminsoity evolution favors a coalesce explosion of a 34\ M$_{\odot}$ Wolf-Rayet star with a 15\,M$_{\odot}$ black hole (BH), demonstrating that some FBOTs may be associated with TDE of a stellar blackhole. 
\end{abstract}

\noindent
Discovery of FBOTs like AT 2018cow reveals a kind of high-energy explosion that likely connects gamma-ray bursts (GRBs) and typical core-collapse supernovae (CCSNe). Unlike normal CCSNe that evolve on timescales of a few weeks to months, the 18cow-like events evolve at a much faster pace, with a typical rise time of only about 3--4 days. Moreover, they are characterized by persistently blue optical colors, subrelativistic outflows or weak jets propagating into dense ambient medium as inferred from radio observations, and X-ray emission that is far in excess of extrapolations from both radio and optical data. Note that some hydrogen-poor transients such as ultrastripped and Type Ibn supernovae (USSNe and SNe Ibn, respectively) are also found to show a fast rise and blue colors at early times, which has been attributed to shock-cooling emission (SCE) from an extended helium-rich envelope\cite{2018A&A...609A.136T, 2020ApJ...900...46Y, 2022ApJ...926..125P, 2023ApJ...959L..32Y}. However, the systems of USSNe and SNe Ibn do not show prominent X-ray and radio emission, contrary to the 18cow-like FBOTs. 

Although a few high-cadence optical surveys over the past decade put efforts into discovering fast-evolving transients, only a small number of AT 2018cow-like luminous FBOTs with associated emission at X-ray and radio wavelengths have been identified, including AT 2020mrf \cite{Yao2022ApJ}, AT 2018lug\cite{2020ApJ...895...49H}, AT 2020xnd \cite{2021MNRAS.508.5138P, 2022ApJ...932..116H, 2022ApJ...926..112B}, 
CSS161010\cite{2020ApJ...895L..23C}, AT 2023fhn\cite{2024MNRAS.527L..47C,2024A&A...691A.329C}, and AT 2022tsd\cite{Ho2023Nature}, suggesting that they are intrinsically rare in the universe. Based on SN classification projects of the Zwicky Transient Facility (ZTF)\cite{2019PASP..131a8002B}, the rate of exotic transients similar to AT 2018cow is at most 0.1\% of the local CCSN rate\cite{2022ApJ...938...85H}. In contrast, the well-known fast-evolving SNe such as SNe Ibn account for about 2.1\% of CCSNe in the local universe\cite{2025A&A...698A.305M}, indicating that the 18cow-like transients might have a different physical origin than the fast-evolving SNe. 

\begin{figure}[htbp]
\centering
\begin{minipage}[t][-0cm][b]{0.48\textwidth}
    \centering
    \begin{minipage}[b]{\textwidth}
        \centering
        \begin{overpic}[width=1.1\textwidth]{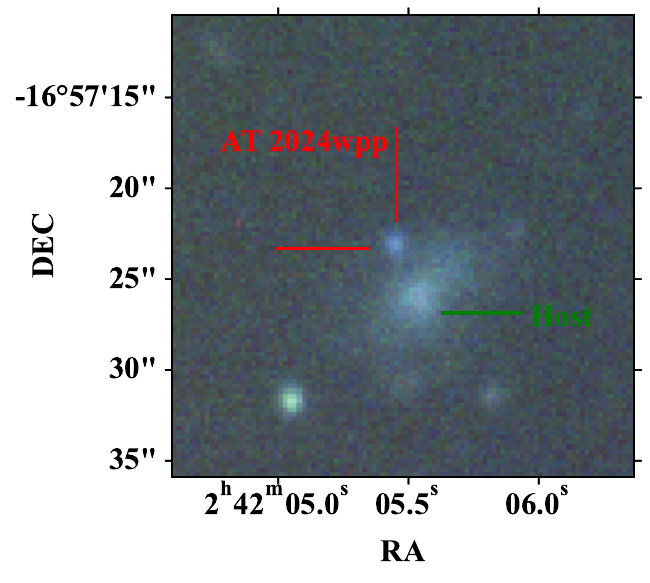}
        \put(-6,66){{\textbf{a}}}
        \end{overpic}
    \end{minipage}
\end{minipage}
\begin{minipage}[b]{0.5\textwidth}
    \centering
    \begin{minipage}[b]{\textwidth}
        \centering
        \begin{overpic}[width=0.75\textwidth]{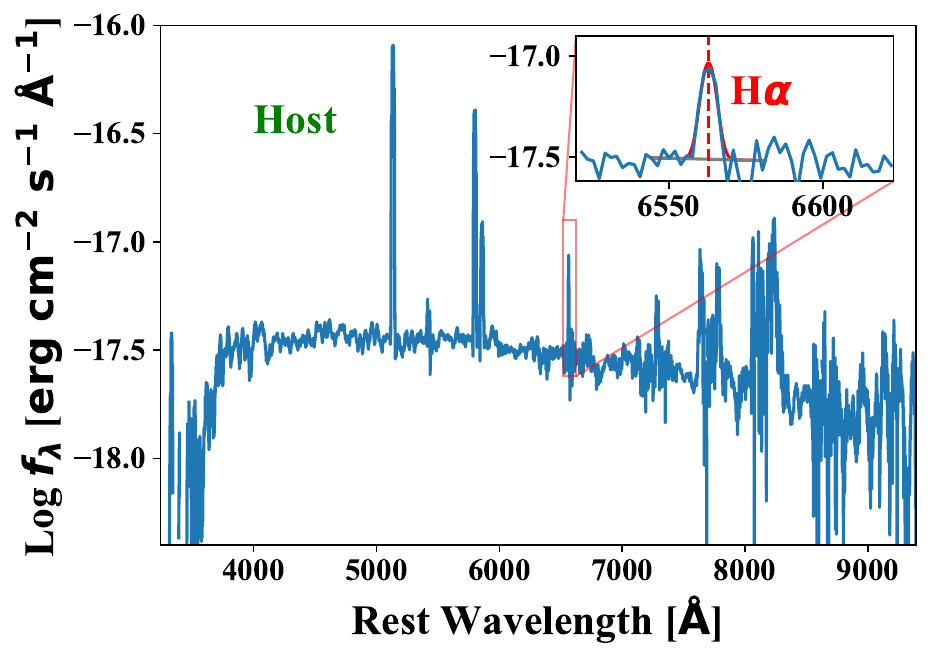}
        \put(-6,60){\textbf{b}}
        \end{overpic}
    \end{minipage}
    \begin{minipage}[b]{\textwidth}
        \centering
        \begin{overpic}[width=0.75\textwidth]{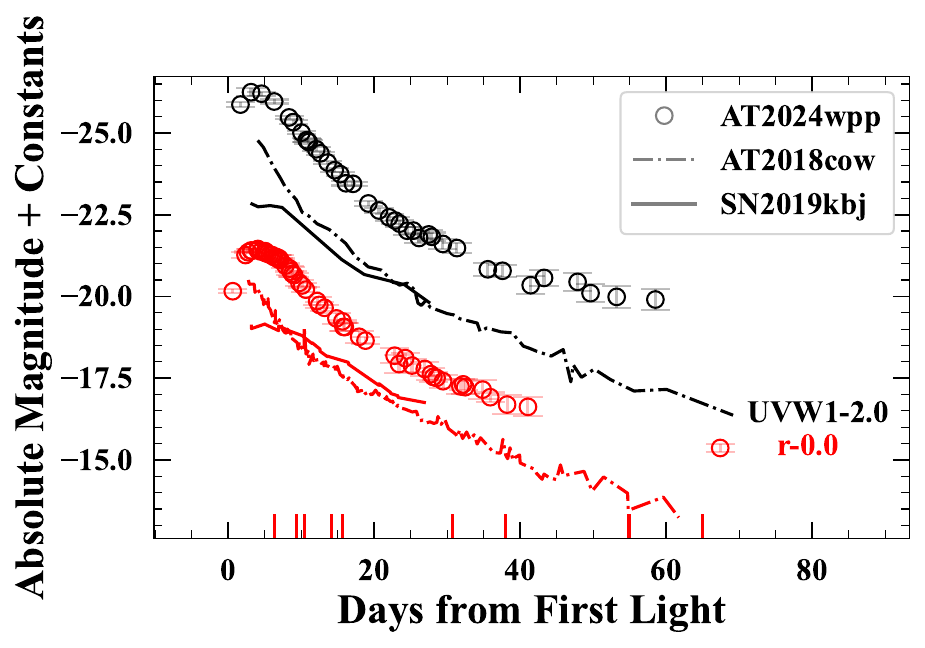}
        \put(-6,60){\textbf{c}}
        \end{overpic}
    \end{minipage}
\end{minipage}
\caption{
\noindent\textbf{AT 2024wpp and its host galaxy at $z = 0.0862$.} 
(a) A Sloan $gri$-band composite image of AT 2024wpp and its host, obtained on 7 Dec. 2024 using the the 10.4~m Gran Telescopio Canarias. The coordinate axes are equatorial coordinates (J2000).
(b) The host-galaxy spectrum taken with GTC on 29 Oct. 2024, which has been corrected with a redshift of 0.0862 estimated using a Gaussian fit to the H$\alpha$ emission line. The inset panel shows the zoomed-in region near the H$\alpha$ line whose rest-frame wavelength is indicated by a red dashed line, together with the redshift-corrected Gaussian profile in red. 
(c) The ultraviolet $UVW1$ and optical $r$ light curves of AT 2024wpp and the comparison objects, including Type Ibn SN 2019kbj\cite{2023ApJ...946...30B} and AT~2018cow\cite{2018ApJ...865L...3P,2019MNRAS.484.1031P,2021ApJ...910...42X}. The light curve of each object has been corrected for reddening and transformed to absolute magnitudes. The vertical red lines represent the phases of spectral observations.
}
\label{fig:AT2024wpp}
\end{figure}

AT 2024wpp, discovered by ZTF on Sept. 26.37 2024 UTC\cite{2024TNSTR3719....1S} (UTC dates are used throughout this paper), represents one of the few known 18cow-like FBOTs. It exploded in a faint dwarf galaxy at redshift $z = 0.0862$, as shown in Figure~\ref{fig:AT2024wpp}. Its classification as a peculiar FBOT was well established by follow-up multiband observations, including a blue featureless spectrum obtained $\sim 5$ days after discovery\cite{2024TNSCR3835....1G} and detections of prominent signals at radio and X-ray wavelengths\cite{2024TNSAN.276....1S, 2024TNSAN.314....1S}. Follow-up multiband photometry spanning ultraviolet (UV), optical, and near-infrared (NIR) bands reveal an extremely energetic explosion\cite{Ofek2025ApJ, 2025arXiv250900951L,2025ApJ...993L...6N,Perley2026arXiv}. 
We started an observational campaign for this peculiar transient at optical, X-ray, and radio wavelengths, covering the phases from a few days to $\sim 110$ days after the explosion. In addition, this object was  extensively monitored by the {\it Swift} UVOT, resulting in well-sampled UV light curves within $\sim 2$ months after explosion (\ref{fig:all_lc}). The $UVW1$- and $r/R$-band light curves of AT 2024wpp respectively show absolute peak magnitudes of $-$24.0 mag and $-$21.5 mag (see Figure~\ref{fig:AT2024wpp}), much brighter than those of stripped-envelope SNe such as 
SN 2019kbj (Type Ibn; \cite{2023ApJ...946...30B}). AT 2024wpp is twice as luminous as AT 2018cow in both UV and optical bands, although both of these  FBOTs show similarly fast evolution around the peak. 

\begin{figure}[htbp]
\centering
\begin{minipage}[b]{0.5\textwidth}
    \centering
        \begin{overpic}[width=0.98\textwidth]{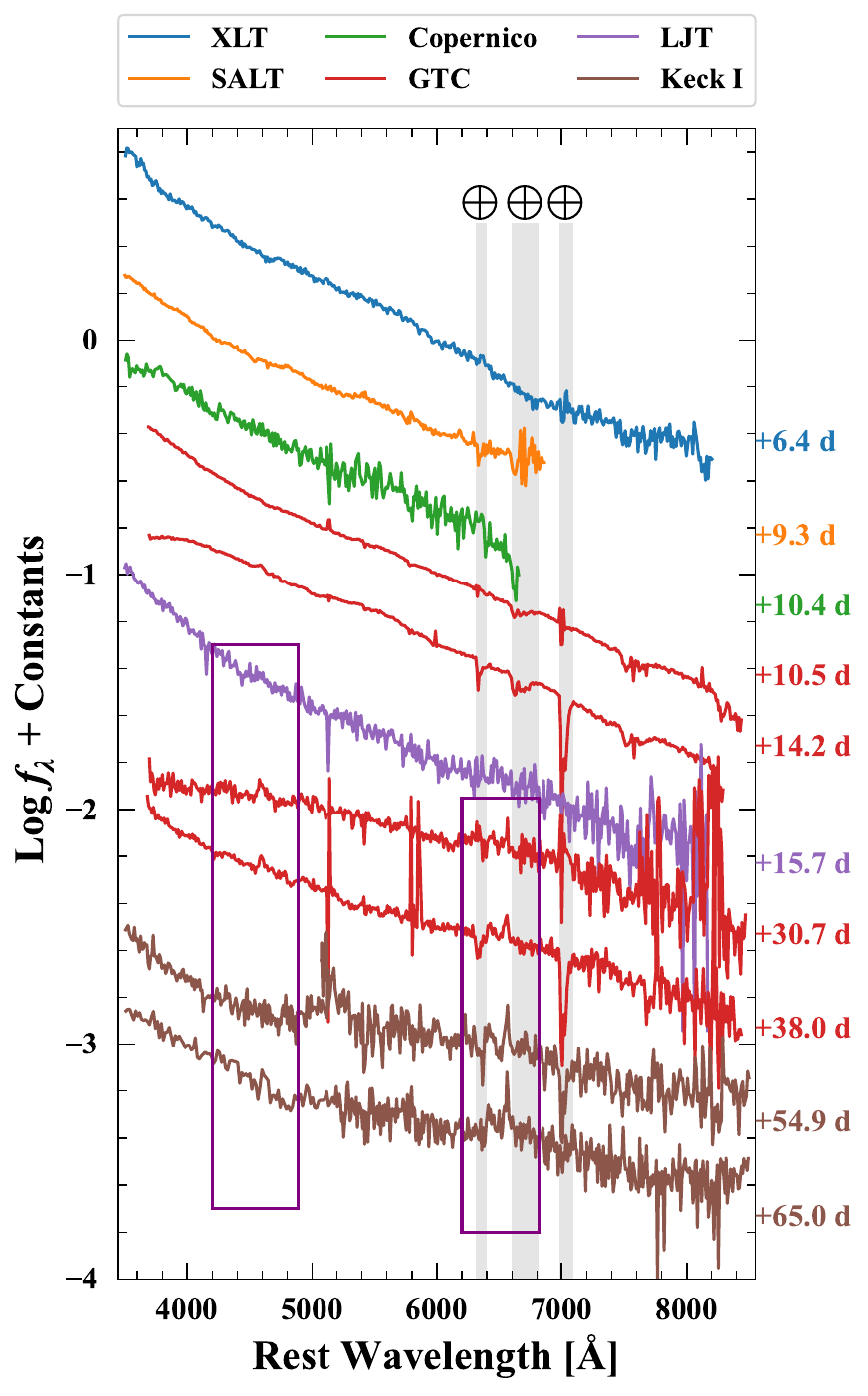}
        \put(0,90){\textbf{a}}
        \end{overpic}
\end{minipage}
\begin{minipage}[t][-0cm][b]{0.48\textwidth}
    \centering
    \begin{minipage}[b]{\textwidth}
        \centering
        \begin{overpic}[width=0.85\textwidth]{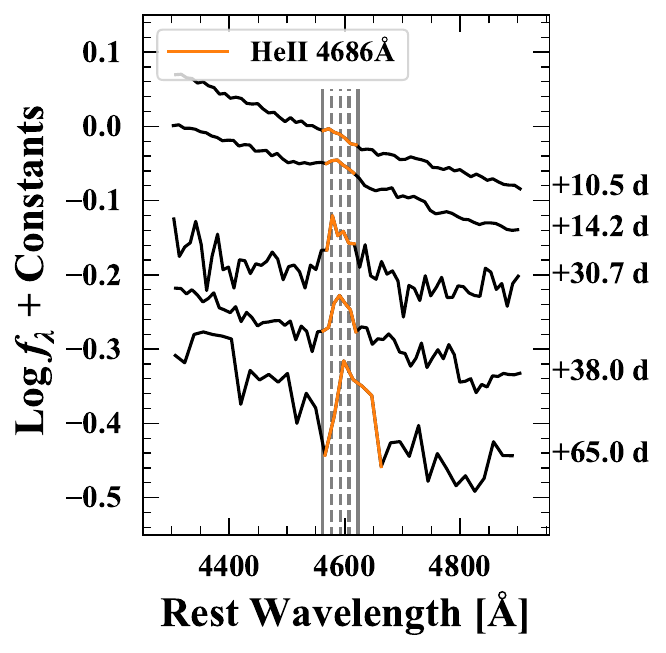}
        \put(-6,66){{\textbf{b}}}
        \end{overpic}
    \end{minipage}
    \begin{minipage}[b]{\textwidth}
        \centering
        \begin{overpic}[width=0.85\textwidth]{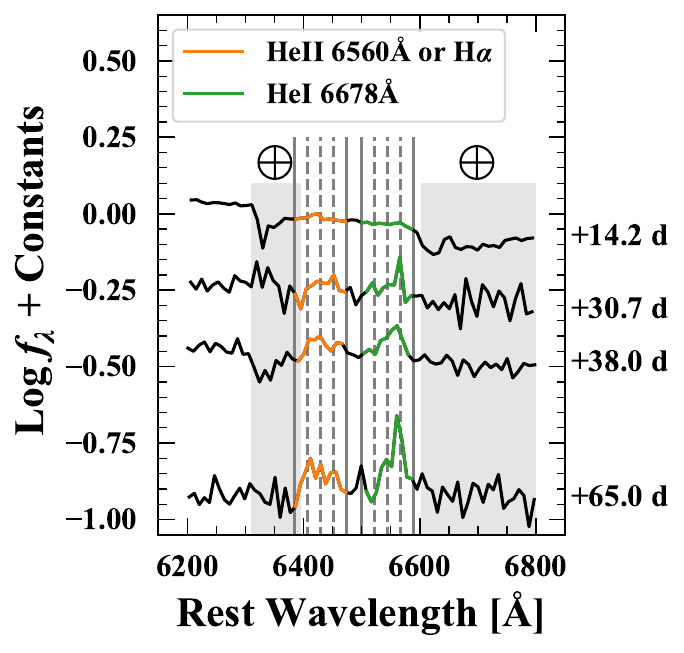}
        \put(-6,68){\textbf{c}}
        \end{overpic}
    \end{minipage}
\end{minipage}
\caption{
\noindent\textbf{Optical spectral evolution of AT~2024wpp.} The spectra are rebinned to 10 \AA\ to increase the S/N. (a) The optical spectra of AT~2024wpp spanning phases from $t\approx +6.4$ days to +65.0 days after the explosion. All spectra have been corrected for reddening and host-galaxy redshift. Spectra taken with different telescopes are denoted in different colors as indicated at the top legend. Telluric absorption lines, visible in some of the spectra, are marked with an ``Earth" symbol and grey vertical bands.  (b) and (c): Zoomed-in regions indicated by the purple frames in panel (a). The potential line regions are highlighted. The vertical lines mark the wavelength positions of lines indicated in the labels, corresponding to velocities from $-$8000 to $-$4000 km s$^{-1}$ with an interval of 1000 km s$^{-1}$. The phase of each spectrum is shown on the right side. 
}
\label{fig:all_spec}
\end{figure}

The spectroscopic behaviour exhibited by AT 2024wpp is also unusual for an SN or any known transient. The earliest spectra in the sequence (see Figure~\ref{fig:all_spec}), sampling around the luminosity peak, exhibit a hot and smooth continuum but possibly with a very broad 
bump feature around 5000--6000 \AA\ (\ref{fig:He5876}).
In the subsequent spectra, a few intermediate-width emission features emerge, including one around 4580 \AA\ that appeared from $t \approx 10$ days and two  at $\sim$6450 and 6550 \AA\ that are visible in later spectra. The $\sim$4580 \AA\ feature could be due to He II $\lambda$4686, while the 6450 \AA\ and 6550 \AA\ features can be identified as He II $\lambda$6560 or H${\alpha}$ and He I $\lambda$6678, respectively. All three emission features are found to have line widths of $\sim 3000$ km s$^{-1}$ and are blueshifted by $\sim 6000$ km s$^{-1}$, indicating that they are formed in the outflow region. About 2 months after the explosion, the spectra are still very blue at short wavelengths, consistent with the high temperature inferred from the spectral energy distribution (SED) obtained at similar phases. Note that a single Planck function cannot give a reasonable fit to the spectra, as a flux excess is found in the NIR regime (and slightly in the $i$ band) (\ref{fig:SED_fit}).

\begin{figure}
\centering
\includegraphics[width=0.7\textwidth]{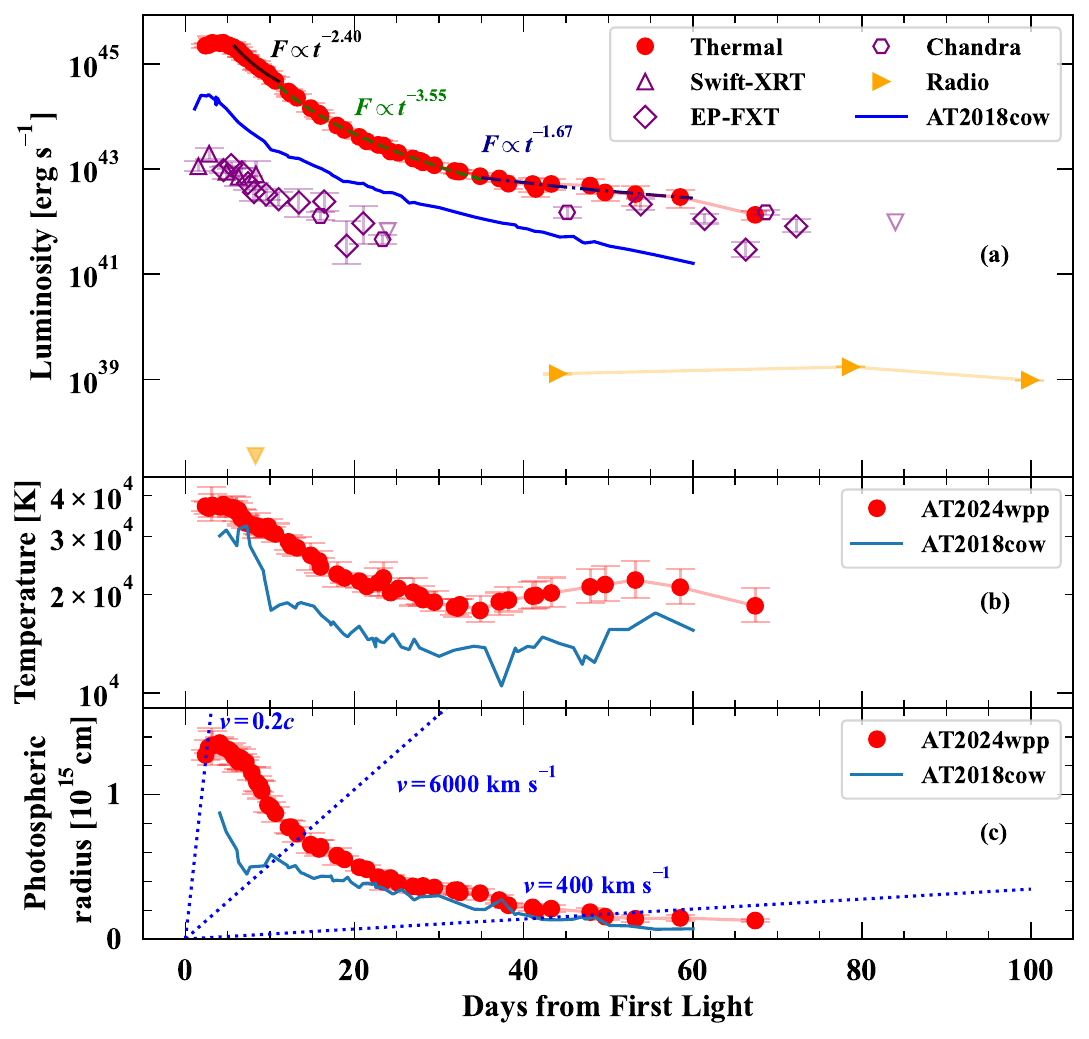}
\caption{
\noindent\textbf{The X-ray, thermal, and radio luminosity evolution of AT~2024wpp}.
{\it Top:} The luminosity evolution in thermal, X-ray, and radio bands as well as an overall comparison of blackbody evolution with that of AT~2018cow\cite{2021ApJ...910...42X}. Open and filled inverted triangles show the upper limits of {\it EP}-FXT and radio luminosities, respectively. Power-law fits to the bolometric light curve at three epochs, $+5.8$ to $+11$ days, $+11$ to $+35$ days, and $+35$ to $+60$ days, are overplotted as a black solid line, a green dashed line, and a dark-blue dash-dotted line, respectively. {\it Middle:} Temperature evolution and the comparison with AT~2018cow. 
{\it Bottom:} Photospheric radius evolution. Blue dotted lines show variation of photosphere size given different assumed velocities.
}
\label{fig:temp_bol}
\end{figure}

Figure~\ref{fig:temp_bol}a shows the evolution of the thermal X-ray and radio  luminosity collected within 70 days post-explosion. Based on the SED constructed by our optical observations and {\it Swift} data, the peak luminosity of the thermal component is $\sim 2.3 \times 10^{45}$ erg s$^{-1}$, an order of magnitude higher than that obtained for AT 2018cow ($2.8 \times 10^{44}$ erg s$^{-1}$) at similar phases, thus being the most luminous known FBOT. After the peak, the bolometric luminosity of AT 2024wpp is found to decay in a power-law mode ($F \propto t^{\beta}$), and the decay rate becomes smaller at late times. 
The decay index $\beta$ is determined to be $-2.40 \pm 0.16$ at phase $t_{1}$(5.8--11 days), $-3.54\pm0.07$ at phase $t_{2}$($\sim$11--35 days), and $-1.67\pm0.61$ at phase $t_{3}$ ($\sim$35--60 days). Although the luminosity evolution shows a rapid decline at early phases, it slows down at $t_{3}$ with a decay rate well consistent with the  classical value of $-$5/3 expected for tidal disruption events (TDEs), suggesting that it could be powered by fallback accretion of additional material onto a BH at this phase. 

Around the time of peak brightness, the thermal component of AT 2024wpp has an unusually high blackbody temperature, $T_{\rm bb} \approx 37,500$ K, much higher than that inferred for any known transient including AT 2018cow. Its post-peak temperature declines rapidly until $t \approx 20$ days since the explosion and appears to show a bump of above 22,000 K at $t \approx 50$ days (see Figure~\ref{fig:temp_bol}, middle panel), a feature also seen in AT 2018cow but at a lower temperature. 


Thanks to the rapid multiband follow-up observations, the earliest SED was available at only $\sim$2.4 days after the explosion, when the corresponding emitting region has a radius as large as $\sim 1.17 \times 10^{15}$ cm. The expansion velocity inferred for the photosphere is thus over 0.18 times the speed of light ($c$), consistent with the broad absorption feature seen in the early-time spectra. 
By $t \approx 4.0$ days, the photospheric radius reached its maximum value ($\sim 1.25 \times 10^{15}$ cm); thereafter, it declines continuously throughout the entire observations (see Figure~\ref{fig:temp_bol}, bottom panel). Such rapid photospheric evolution is atypical for an SN but is seen in the prototype FBOT AT 2018cow, further justifying the classification of AT 2024wpp as an 18cow-like event.

Radio observations of AT 2024wpp were obtained at several epochs with the Australian Telescope Compact Array (ATCA) under project code CX582 (see Method 2.8). Data were obtained at lower central frequencies (5.5 GHz and 9.0 GHz) relative to the reference \cite{2025ApJ...993L...6N}, which allows the construction of a radio SED with wider frequency coverage. As shown in \ref{fig:radio_sed}, the combined SED established during $t \approx 40$--110 days since the explosion can be fit with a broken power law by assuming a synchrotron self-absorption (SSA) spectrum \cite{1998ApJ...499..810C}. At $t \approx 45$--50 days, the combined SED is found to show large scatter near 9--10 GHz, perhaps due to a fast variability timescale at this frequency. The SED at $t \approx 80$--110 days provides a better fit, with the parameters of shock velocity, shock radius, and density of the ambient medium being determined as $v_{\rm sh}> 0.2$--0.3$c$, $R \approx (6$--$7) \times 10^{16}$ cm, and $n_{e} \approx 120$--230 cm$^{-3}$ (see Method 2.8 and \ref{tab1}). 

The radio light curve of AT 2024wpp shows a similar luminosity evolution as other FBOTs (see \ref{fig:radio}), reaching a flux peak at $t \approx 80$ days. At the central frequency of 9 GHz, it achieves a peak luminosity $> 2 \times 10^{39}$ erg s$^{-1}$, noticeably higher than normal CCSNe and those that exploded in dense circumstellar material (CSM). The radio luminosity of AT 2024wpp lies between that of the prototype FBOT AT 2018cow and the most luminous event (see \ref{fig:radio}). Stronger radio luminosity is consistent with the inference of a relativistic shock and the accompanied large amount of energy swept into the ambient medium, (1.2--$2.2) \times 10^{49}$ erg.   

With the Follow-up X-ray telescope (FXT) onboard the {\it Einstein Probe (EP)} satellite, we triggered the X-ray observations of AT 2024wpp spanning the phase from $t \approx 3.5$ days to $\sim$90 days after the explosion. An X-ray source is clearly detected at the position of the optical transient, and the complete X-ray light curve in the 0.5--10 keV band is shown in Figure~\ref{fig:temp_bol}, top panel. 
Our X-ray data are complemented by critical early-time and late-time observations from the {\it Swift} X-ray Telescope ({\it Swift}-XRT)\cite{Srinivasaragavan2024TNS} and the {\it Chandra X-ray Observatory} (CXO)\cite{Margutti2024TNS}, which helps establish three distinct phases of evolution: initial rise, subsequent power-decay, and late-time brightening (see also  \ref{fig:x ray light curve}). AT 2024wpp rises to a peak of 2$\times$10$^{43}$ erg s$^{-1}$ at 2-3 days after the explosion, and it then enters the power-law decay phase until t$\sim$1 month. After that, the X-ray luminosity is found to re-brighten and reach a second peak of 2$\times$10$^{42}$ erg s$^{-1}$ at t$\sim$2 months, and the transient became no longer detected by $t \approx 3$ months. Such an X-ray luminosity evolution differs significantly from that of AT 2018cow and other known FBOTs. Notably, the X-ray rebrightening behavior seen in AT 2024wpp is consistent with its optical evolution observed at similar phase (see Figure 3), suggesting the emergence of a new energy component.


The X-ray light curve of AT 2024wpp appears to exhibit X-ray variability on a timescale less than one day, although the measurement errors are relatively large and the data are not densely sampled as for AT 2018cow. 
Such a complex X-ray evolution cannot be due to interaction of SN ejecta with CSM; instead, it likely suggests the existence of a central engine. Moreover, in contrast to AT 2018cow and other FBOTs, the X-ray spectrum $F_{\nu}\sim\nu^{-\beta}$ of AT 2024wpp is found to show peculiar evolution, with the spectral index varying from $-$0.97 at t$\sim$ phase to +1.33 at t$\sim$50 day (see \ref{fig:XraySpectrum}). The appearance of an inverted spectrum  around the rebrightening phase thus favors for the presence of a new emission component at late time. 
This is also demonstrated by the emergence of 
intermediate-width emission features due to blueshifted He~II and He~I lines in the spectra. 
It is important to note that the X-ray emission observed from AT 2024wpp cannot be described as an extension of the radio synchrotron spectrum, nor by inverse Compton scattering of UV-through-NIR photons by electrons accelerated in the forward shock. The X-ray flux exceeds the radio synchrotron extrapolation, further confirming a distinct central energy source for the late rebrightening.

Different scenarios for the power source have been applied to scrutinize both the thermal and X-ray luminosity evolution of AT 2024wpp\cite{Omand2026arXiv}. As a classical power source for SNe, the radioactive Ni$^{56}$/Co$^{56}$-driven model cannot produce the observed optical luminosity of this transient with a fast rise to the extraordinarily high peak, as shown in \ref{fig: model_sim}. While a hybrid model provides a better fit by adding a new component from ejecta interaction with CSM, it hardly explains the steep decay during the final phase.

The magnetar model offers more compelling explanations for the robust profile of the luminosity evolution, especially the early-time light curves (\ref{fig: model_sim}). However, it cannot produce as much light as the observations at late phases. To address this discrepancy, an accreting magnetar model is developed to incorporate the contribution of late-time fallback accretion onto the newborn magnetar (see Method 2.9). In this model, a newly formed magnetar, with an initial spin period of only 4.6 ms and an intense surface dipole magnetic field of $6\times10^{14}$ G, powers the SN ejecta with a mass of 0.66 M$_\odot$ at a speed of 0.14$c$, well reproducing the observed early-time light curves in optical and X-ray bands. An accretion process started from $t \approx 41$ days with an initial rate of $1.4 \times 10^{-6}$ M$_\odot$ s$^{-1}$, which provides an extra power source to boost the late-phase luminosity (\ref{fig:models_m}).

\begin{figure}
\centering

\begin{minipage}[b]{0.75\textwidth}
    \centering
        \begin{overpic}[width=1\textwidth]{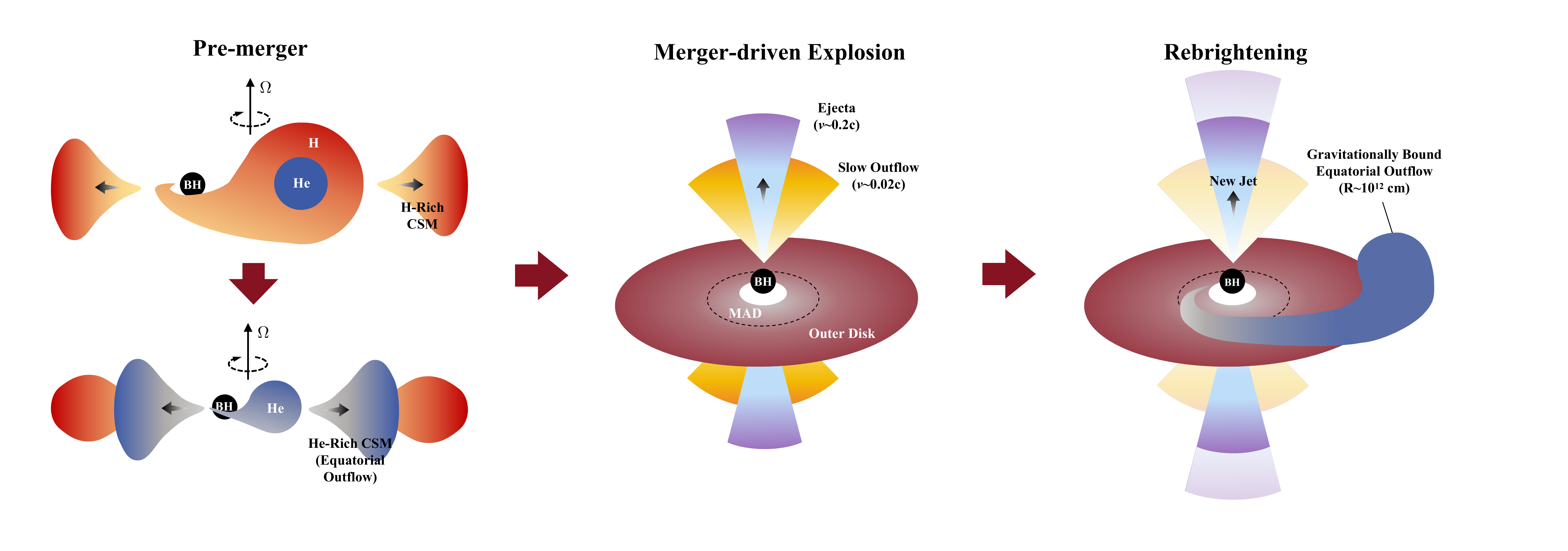}
        \put(0,35){\textbf{a}}
        \end{overpic}
\end{minipage}
\begin{minipage}[b]{0.62\textwidth}
    \centering
        \begin{overpic}[width=1\textwidth]{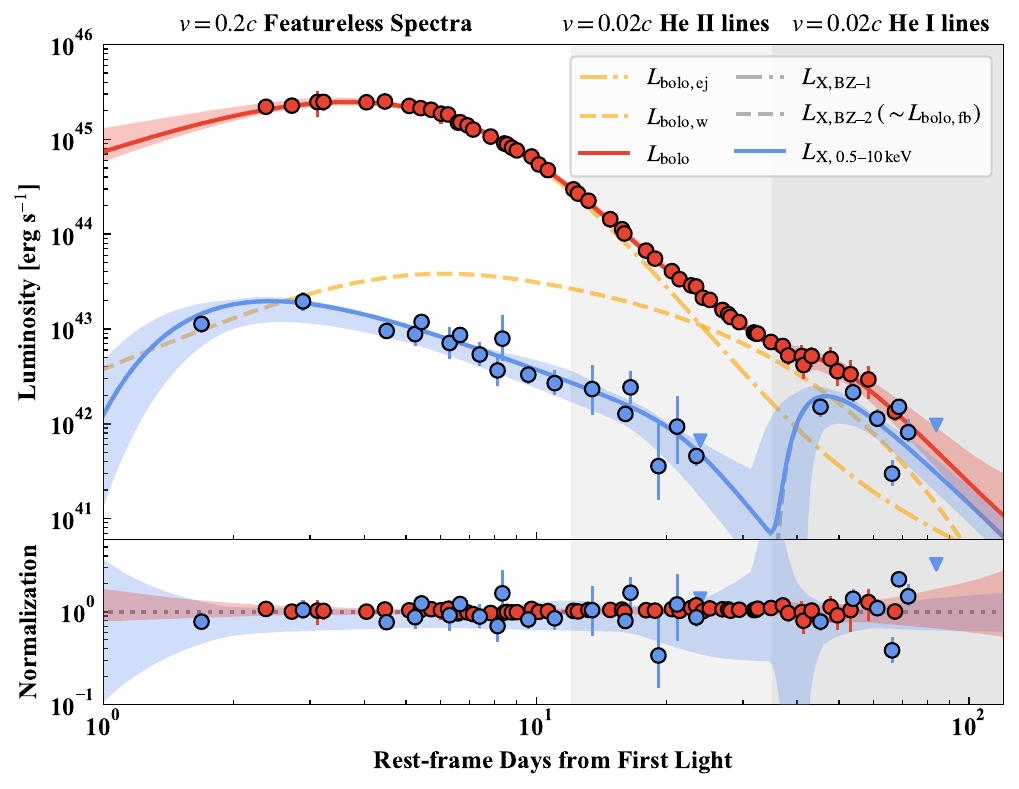}
        \put(0,72){\textbf{b}} 
        \end{overpic}
\end{minipage}

\caption{
\noindent\textbf{The WR-BH merger-driven explosion scenario invoked for AT 2024wpp.
}
    (a) The WR-BH merger scenario\cite{Metzger2022ApJ, Klencki2025arXiv}, in which the WR companion is disrupted and accreted onto the BH, resulting in a luminous explosion.  
    At t$\gtrsim 30$ days post explosion, a portion of gravitationally bound equatorial outflow falls to the BH\cite{Pejcha2016MNRAS}, leading to the X-ray rebrightening and flattening in optical. 
    (b) {\it Upper panel:} The red and blue lines represent the best-fit models for the bolometric and X-ray light curves, respectively. The model consists of several emission components: the bolometric luminosity includes contributions from the initial shock interaction between the low-mass ejecta and the He-rich CSM, supplemented by the reprocessing of X-rays ($L_{\rm bolo, \, ej}$; orange dot-dashed line), the interaction between the CSM and the nonpolar slow outflow ($L_{\rm bolo, \, w}$; orange dotted line), and the emission from the second BZ jet caused by fallback accretion ($L_{\rm bolo, \, fb}$; grey dotted line).
    The X-ray emission includes contributions from the primary ($L_{\rm X, \, BZ–1}$; grey dot-dashed line) and second ($L_{\rm X, \, BZ–2}$; grey dotted line) jets powered by the BZ mechanism. 
    {\it Lower panel:} The fit residuals, expressed as the ratio of the observed data to the best-fit model values. 
    }
\label{fig:model fit}
\end{figure}

Alternatively, accretion onto a BH is a competitive candidate for the energy source behind the extremely high optical luminosity (\ref{fig: model_sim}), though an additional mechanism is required to account for the late-time bump. Note that a delayed-merger scenario between a WR star and a BH was proposed in Ref. \cite{Metzger2022ApJ}, where the WR/BH binary first undergoes a common-envelope phase and loses the hydrogen-rich envelope (see Figure~\ref{fig:model fit}a). Subsequently, 
TDEs
of the helium core leads to an equatorial outflow from the Lagrange L2 point, and eventually it merges on a viscous delayed timescale. The merger-powered central engine is expected to produce luminous multiband emission. However, this model fails to match the detailed X-ray luminosity evolution. 
Moreover, the assumption that the optical emission is entirely powered by X-ray reprocessing provides a poor fit to the optical light curve. Also, the rapid ($<0.5$ day) X-ray fluctuations are more naturally interpreted as direct radiation from the central engine. 

Owing to these limitations, we modify this model to allow the application to AT 2024wpp. Accretion onto a rapidly spinning BH forms a magnetically arrested disk (MAD)\cite{Igumenshchev2003ApJ, Igumenshchev2008ApJ, Tchekhovskoy2015MNRAS, You2023Sci} and powers an anisotropic outflow via the Blandford-Znajek (BZ) mechanism\cite{Blandford1977MNRAS}, producing the stable X-ray emission, while optical emission arises from shock interaction and X-ray reprocessing (see Methods for details).
Our fitting results reveal a scenario that is consistent with the observations: a merger occurs between a $\sim 34$ M$_{\odot}$ WR star and a $\sim 15$ M$_{\odot}$ BH.
The BZ mechanism-powered jet consists of the low-mass ejecta ($\sim 0.3$ M$_{\odot}$) with a velocity of $\sim 0.2c$ and a slow outflow with a velocity of $\sim 0.02c$. 
At late times, i.e., $\gtrsim 30$ days post explosion, a portion of the gravitationally bound equatorial outflow\cite{Pejcha2016MNRAS} at $R\approx 10^{12}\,{\rm cm}$ falls into the BH with an inflow rate of $\sim 0.1$ M$_{\odot} {\rm yr}^{-1}$, initiating the secondary accretion. This delayed-fallback process explains the X-ray rebrightening and the optical flattening at the same time, unprecedented features in FBOTs. 
Furthermore, our model is found to be consistent with three distinct spectroscopic evolution phases (see Figure~\ref{fig:model fit}b): a featureless continuum with very broad absorption in the early stage ($t<12$ days), the emergence of He II $\lambda$4686 emission  blueshifted at $v \approx 0.02c$ ($t>12$ days), and the emergence of He I $\lambda$6678 emission in the late-time rebrightening phase ($t>35$ days).


Thus, the WR-BH merger model (with double accretion phases) provides the best fit to the multiband luminosity evolution of AT 2024wpp, indicative of the mechanism of fallback accretion onto a central engine as the main power source enlightening this luminous transient. This model advances the FBOT physics, linking such kind of exotic transients to compact object mergers and constraining evolution pathways of massive binary stars.


\clearpage 

%
\bibliography{sn-bibliography} 
\bibliographystyle{sciencemag}

%
%
%
%
%
%


\section*{Acknowledgments}
\paragraph*{Funding:}
We are grateful for the support of the staff of the 10.4~m Gran Telescopio Canarias (GTC), Keck~I 10~m telescope, Xinglong 2.16~m telescope, Lijiang 2.4~m telescope (LJT), SALT, and {\it Swift}/UVOT. 
The W. M. Keck Observatory is operated as a scientific partnership among the
California Institute of Technology, the University of California, and
NASA; the observatory was made possible by the generous financial
support of the W. M. Keck Foundation. This research was based partially on observations made with the Gran Telescopio Canarias (GTC), (Program GTCMULTIPLE2B-24B) installed at the Spanish Observatorio del Roque de los Muchachos of the Instituto de Astrofísica de Canarias, on the island of La Palma.
This work was partially based on observations collected at Copernico and Schmidt telescopes (Asiago Mount Ekar, Italy) of the INAF -- Osservatorio Astronomico di Padova.

The work of X.-F.W. is supported by the National Natural Science Foundation of China (NSFC grants 12288102, 12033003, and 11633002), the Ma Huateng Foundation, and the Tencent Xplorer Prize. 
N.E.R. acknowledges support from the PRIN-INAF 2022, ``Shedding light on the nature of gap transients: from the observations to the models'' and from the Spanish Ministerio de Ciencia e Innovaci\'on (MCIN) and the Agencia Estatal de Investigaci\'on (AEI) 10.13039/501100011033 under the program Unidad de Excelencia Mar\'ia de Maeztu CEX2020-001058-M.
The work of Xuefeng Wu is supported by 
NSFC grant 12321003.
J.C. acknowledges funding by the Australian Research Council (ARC) Discovery Project DP200102102, and the ARC Centre of Excellence for Gravitational Wave Discovery (OzGrav) through project numbers CE170100004 and CE230100016.
Y.-Z. Cai is supported by 
NSFC grant 12303054, the National Key Research and Development Program of China (grant  2024YFA1611603), and the Yunnan Fundamental Research Projects (grants  202401AU070063 and 202501AS070078).

A.V.F.'s research group at U.C. Berkeley was supported by the Christopher R. Redlich Fund, Gary and Cynthia Bengier, Clark and Sharon Winslow, Alan Eustace and Kathy Kwan (W.Z. is a Bengier-Winslow-Eustace Specialist in Astronomy), Timothy and Melissa Draper, Briggs and Kathleen Wood, Ellen and Alan Seelenfreund (T.G.B. is Draper-Wood-Seelenfreund Specialist in Astronomy), and numerous other donors.
A.V.F. is grateful for the hospitality of the Hagler  
Institute for Advanced Study as well as the Department of Physics and 
Astronomy at Texas A\&M University during part of this investigation.


\paragraph*{Competing interests:}
There are no competing interests to declare.
\paragraph*{Data and materials availability:}
\textbf{Supplementary Information} is available for this paper.\\
{\bf Correspondence and requests for materials} should be addressed to Xiaofeng Wang (\url{wang_xf@mail.tsinghua.edu.cn}).


\subsection*{Supplementary materials}
Materials and Methods\\
Figs. S1 to S10\\
Tables S1 to S9\\
References \textit{(42-\arabic{enumiv})}\\ 
Data S1


\newpage


\renewcommand{\thefigure}{S\arabic{figure}}
\renewcommand{\thetable}{S\arabic{table}}
\renewcommand{\theequation}{S\arabic{equation}}
\renewcommand{\thepage}{S\arabic{page}}
\setcounter{figure}{0}
\setcounter{table}{0}
\setcounter{equation}{0}
\setcounter{page}{1} 


\begin{center}
\section*{Supplementary Materials for\\ \scititle}

Jialian Liu, 
Bao Wang, 
Xiaofeng Wang, 
David Aguado,
Weili Lin,\\
Nancy-Elias Rosa, 
Qichun Liu, 
Frederick Poidevin,
Ismael Perez-Fournon,\\ 
Long Li, 
Ailing Wang,
Yi Yang, 
Zigao Dai, 
Alexei V. Filippenko,\\ 
Thomas G. Brink, 
Di Xiao, 
Wenxiong Li, 
Yifang Liang,
Xuefeng Wu,\\ 
Samaporn Tinyanont,
Jinjun Geng, 
Shengyu Yan, 
Weimin Yuan,
Jujia Zhang,\\ 
Xiangyun Zeng, 
WeiKang Zheng, 
Yuanming Wang,
Tao An, 
YongZhi Cai,\\
Jeff Cooke, 
Lixin Dai, 
Andrea Farina, 
Maokai Hu,
Ye Li,
Chichuan Jin,\\ 
Yuan Liu, 
David Lopez Fernandez-Nespral, 
Alicia Lopez Oramas,\\ 
Andrea Reguitti,
Xinwen Shu, 
Cuiying Song, 
Hui Sun,\\ 
Ning-chen Sun, 
Lifan Wang, 
Tinggui Wang, 
Junjie Wei,
Qingyu Wu,\\
Danfeng Xiang,
Lei Yang,
Liping Li,
and Zhenyu Wang\\
\end{center}

\subsubsection*{This PDF file includes:}
Materials and Methods\\
Figures S1 to S10\\
Tables S1 to S9\\
Captions for Data S1

\subsubsection*{Other Supplementary Materials for this manuscript:}
Data S1

\newpage


\subsection*{Materials and Methods}

\subsubsection*{Host redshift and extinction}~\label{sec:host}

We obtained a host-galaxy spectrum of AT 2024wpp on UTC 29 Oct. 2024 with OSIRIS mounted on the 10.4~m Gran Telescopio CANARIAS (GTC), as shown in \ref{fig:AT2024wpp}. Using a Gaussian profile to fit the H$\alpha$ emission, we estimated a redshift of $z=0.0862$, consistent with that reported \cite{2024TNSAN.280....1P}, which is adopted throughout this work. Then the distance modulus is $DM=38.05\pm0.15$ mag, assuming the cosmological parameters of the Lambda cold dark matter ($\Lambda\mathrm{CDM}$) model: H$_0 = 67.4~\mathrm{km~s^{-1}~Mpc^{-1}}$, $\Omega_m = 0.315$, and $\Omega_\Lambda = 0.685$ \cite{2020A&A...641A...6P}. As AT~2024wpp is located in the outskirts of its host, it thus suffered little host extinction, which is consistent with the absence of Na\,{\sc i}\,D in its spectra. The Galactic reddening, namely the total reddening, is taken as $E(B-V)=0.0253$ mag according to the dust map \cite{2011ApJ...737..103S}.

\subsubsection*{Photometry of follow-up observations}~\label{sec:obs_lc}

The photometric follow-up campaign on SN\,2024wpp was carried out by the 0.8\,m Tsinghua University-NAOC\footnote{National Astronomical Observatories of China (NAOC), China.} telescope (TNT)\cite{2008ApJ...675..626W} at Xinglong Station of NAOC (China), the Lijiang 2.4~m telescope \cite{2015RAA....15..918F} (LJT) of Yunnan Astronomical Observatories (China), the AZT-22 1.5 m telescope (AZT) at Maidanak Astronomical Observatory \cite{2018NatAs...2..349E}, the Thai Robotic Telescope (TRT), the telescopes of the Las Cumbres Observatory network (LCO; program IAC2024B-004 - P.I. Frédérick Poidevin), the Schmidt 67/91 cm Telescope (67/91-ST) of the Osservatorio Astronomico di Asiago (Italy), and GTC at the Roque de Los Muchachos Observatory (Spain).

All images were preprocessed following standard routines, including bias subtraction, flat-field correction, dark-current correction, and cosmic-ray removal. Point-spread-function (PSF) photometry was performed with the instrumental magnitudes calibrated using the Gaia Synthetic Photometry Catalogue \cite{2023A&A...674A..33G} except for the GTC images, for which  aperture photometry with small-sized apertures was performed to increase the signal-to-noise ratio (S/N) of the results. To increase the accuracy of the aperture correction, we used two nearby stars that have the closest full width at half-maximum intensity (FWHM) to the SN to get the photometric calibration with the Pan-STARRS Release 1 \cite{2016arXiv161205560C} that includes these two stars. Note that there is almost no difference between the results of the Pan-STARRS Release 1 and the Gaia Synthetic Photometry Catalogue in Sloan filters.

AT~2024wpp was also observed by the Ultraviolet/Optical Telescope (UVOT) \cite{2004ApJ...611.1005G,2005SSRv..120...95R} on the {\it Neil Gehrels Swift Observatory} \cite{2004ApJ...611.1005G} in three UV ($UVW2$, $UVM2$, $UVW1$) and three optical ($U$, $B$, $V$) filters. We extracted {\it Swift} photometry using \texttt{Swift\_host\_subtraction}\footnote{\url{https://github.com/gterreran/Swift_host_subtraction}} with the latest \emph{Swift} calibration database\footnote{\url{https://heasarc.gsfc.nasa.gov/docs/heasarc/caldb/swift/}}. To remove the contamination from the host galaxy, especially at $t > +40$ days after explosion when the SN and the host have a comparable brightness, template subtraction was applied using the latest image, where the SN signal is negligible. In addition, we include the $g$-band photometry of the All-Sky Automated Survey for Supernovae (ASSASN) \cite{2014ApJ...788...48S}, the $L$-band photometry of MJD 60,579.059 from the Gravitational-wave Optical Transient Observer (GOTO), and the $w$-band photometry of MJD 60,579.556 from Pan-STARRS1 (PS1).

\subsubsection*{Parameters of the comparison sample}~\label{sec:lc_compare}

The comparison sample shown in \ref{fig:AT2024wpp} includes AT~2018cow (the first discovered FBOT; \cite{2018ApJ...865L...3P,2019MNRAS.484.1031P,2021ApJ...910...42X}) and SN 2019kbj (an SN Ibn; \cite{2023ApJ...946...30B}). The light curves of all SNe presented here have been corrected for reddening and transformed to absolute magnitudes. 

For AT~2024wpp, the host reddening is negligible and only the Galactic reddening of $E(B-V)=0.0253$ mag is considered. For AT~2018cow, we adopt a reddening of $E(B-V)=0.076$ mag and  DM $=34.10\pm0.15$ mag \cite{2018ApJ...865L...3P}. For SN~2019kbj, we adopt a reddening of $E(B-V)=0.05$ mag and DM $=36.66\pm0.15$ mag \cite{2023ApJ...946...30B}.

\subsubsection*{Spectroscopy}~\label{sec:obs_spec}

Optical spectra of AT~2024wpp were collected with several different instruments, including BFOSC mounted on the Xinglong 2.16~m telescope (XLT) \cite{2016PASP..128j5004Z} at Xinglong Station of NAOC (China), YFOSC on LJT, AFOSC on the 1.82~m Copernico Telescope (Copernico) at the Mount Ekar Observatory (Italy), LRIS \cite{1995PASP..107..375O} on the 10~m Keck-I telescope on Maunakea, RSS on the Southern African Large Telescope (SALT) at South African Astronomical Observatory (South Africa), and OSIRIS on GTC at the Roque de Los Muchachos Observatory (Spain). The Keck~I/LRIS spectrum was reduced using the \texttt{LPipe} pipeline \cite{2019PASP..131h4503P}, while the standard IRAF routine was used to reduce other spectra. Spectrophotometric standard stars were taken on the same nights for flux calibration of the SN spectra. We corrected for atmospheric extinction using the extinction curves of local observatories, and telluric lines were removed from most of the spectra.

Optical spectra of AT~2024wpp are shown in Fig.~\ref{fig:all_spec} and the journal of these observations can be found in \ref{table:spec}. Evolution of the 4600 \AA\ and 6500 \AA\ regions is also presented in Fig.~\ref{fig:all_spec}. For the 4600 \AA\ region, we noticed an emission feature appearing in the $t\approx+14.2$ days spectrum, which is not seen in other FBOTs. Two other emission lines also seem to appear in the 6500 \AA\ region at $t\approx+30.7$ days and become evident at $t\approx+38.0$ days. These lines have a consistent velocity distribution if we attribute them to He\,{\sc ii} 4686 \AA, He\,{\sc ii} 6560 \AA\ or H$\alpha$, and He\,{\sc i} 6678 \AA, respectively. Note that the narrow H$\alpha$ line at the rest wavelength is due to host-galaxy contamination. 

Comparisons of the broad emission feature in the 5000--6000 \AA\ region are presented in Fig.~\ref{fig:He5876}. This feature is clearly seen in the early spectra of AT~2018cow\cite{2021ApJ...910...42X}, 2020mrf\cite{2020TNSCR1846....1B}, and 2024wpp after subtracting a blackbody continuum, but it seems to be absent in the spectrum of AT~2020xnd\cite{2021MNRAS.508.5138P}. It could come from the He\,{\sc i} 5876 \AA\ line with a relativistic velocity\cite{2025MNRAS.537.3298P}. 

\subsubsection*{Bolometric light curve}~\label{sec:Bol_Temp}

To estimate the bolometric light curve of AT~2024wpp, we assumed that it has a blackbody SED shape and used {\it Swift} $UVM2$, $UVW1$, and ground-based $ugri$ photometry to calculate the blackbody parameters. The data were interpolated in each filter so that the bolometric luminosity could be calculated at the same phase based on the LCO $g$-band data. Before MJD 60585.906, the $UVM2$ band was not used owing to its poor sampling. After $t \approx +37$ days, owing to the scarcity of multiband observations, we tried to fit a blackbody to three late-time spectra obtained at $t\approx+38.0$, $+54.9$, and $+65.0$ days to examine the temperature evolution. However, this temperature is much lower than that estimated by the photometry, which is possibly due to nonthermal emission on the red side. We found that the blackbody fit to part of the $t\approx+38.0$ days spectrum ($\sim3700$--4500 \AA) gave a temperature of $T=19,100_{-1500}^{+1800}$ K, which is consistent with that given by the photometry at $t\approx+37.1$ days ($T=19,000_{-1700}^{+1300}$ K). Then we also fit a blackbody to the $\sim3700$--4500 \AA\ regions of the other two nebular spectra and obtained $T=22,400_{-2700}^{+3400}$ K and $T=18,400_{-1600}^{+2000}$ K at $t\approx+54.9$ and $+65.0$ days, respectively. We applied a linear interpolation of these results to estimate the temperature at the epochs when the $g$-band and $UVW1$-band photometry are available\footnote{Note that the last $g$ observation is slightly beyond the phase range of the spectra, and is assumed to have the same temperature as the last spectrum by adding an extra uncertainty of 400 K.}, and then calculated the blackbody luminosity. All the results of the bolometric luminosity and temperature can be found in \ref{table:blackbody}, and the comparisons with AT~2018cow\cite{2021ApJ...910...42X} are shown in Figure~\ref{fig:temp_bol}.

Based on the spectral features and temperature evolution, the luminosity evolution can be divided into three phases: $t_{1} \approx +5.8$--11 days, $t_{2} \approx 11$--35 days, and $t_{3} \approx 35$--60 days. Applying power-law fits ($F \propto t^{-\beta}$) to the bolometric light curve at the above three phases yields the decay index $\beta$ as $2.40 \pm 0.16$, $3.54\pm0.07$, and $1.67\pm0.61$, respectively. 
The inferred late-time slope is sensitive to the fitting window (e.g., $\beta=2.07\pm0.30$ when fitting 30--60\,d). 
For comparison, the corresponding decay index measured for AT~2018cow is $2.30\pm0.04$, $2.57\pm0.01$, and $3.46\pm0.04$, respectively. This indicates that, during $t_{3}$, the luminosity evolution of AT 2024wpp shows a large discrepancy from that of AT 2018cow, while it is consistent with that of TDEs ($\beta = 5/3$). This suggests that the late-time luminosity of AT 2024wpp is related to  fallback accretion onto a compact object.

\subsubsection*{Spectral energy distribution evolution}~\label{sec:SED}

Combined with the near-infrared photometry of Ref. \cite{2025MNRAS.537.3298P,2025arXiv250900951L}, we examined the SED of AT~2024wpp from UV to near-infrared (NIR) bands at $t\approx+2.4$, $9.8$, $18.8$, $28.1$, $37.1$, and $67.4$ days after the explosion. As shown in Figure~\ref{fig:SED_fit}, the SED can be well described by a single blackbody component (see Section~\ref{sec:Bol_Temp}) at least until $t\approx+9.75$ days, while a weak excess in the NIR is found at $t\approx+18.80$ days, and it quickly became strong at $t\approx 28.1$ days. At $t\approx 37.1$ and $67.4$ days, even the $i$-band flux ($\sim 7000$ \AA) shows a weak excess, consistent with the spectral observations at $t\approx 38$ and $65.0$ days. 

\subsubsection*{X-ray observations}~\label{sec:xray}

\subsubsection*{X-ray data reduction}~\label{sec:xray data}


The {\it Einstein Probe (EP)}\cite{Yuan_2022} is a space mission designed for time-domain high-energy astrophysics. The Follow-up X-ray Telescope (FXT)\cite{Chen2021SPIE}, which consists of two coaligned modules (FXT-A and FXT-B), provides precise positioning capabilities of $10''$ in the 0.5--10 keV band, making it ideal for monitoring interesting transients to trace their temporal and spectral evolution.

Our monitoring campaign for AT2024wpp with {\it EP}-FXT began on 2024 Sept. 30 (PIs B. Wang and X. Wang), comprising 17 observations with a total exposure of about 110~ks spanning a period of $\delta t \approx 5.2$--91.6 days post-explosion.
Data reduction was carried out using the Follow-up X-ray Telescope Data Analysis Software (FXTDAS v1.20)\footnote{\url{http://epfxt.ihep.ac.cn/analysis}} together with the corresponding calibration database (CALDB v1.20). Source events were extracted from a circular region centered on the X-ray position, while background events were taken from a nearby source-free region. The corresponding redistribution matrix file (RMF) was taken from CALDB, while the ancillary response file (ARF) was generated with \texttt{fxtarfgen} based on the source extraction region. The 0.5--10 keV spectra were fitted in \texttt{XSPEC} using Cash statistics with an absorbed power-law model (\texttt{tbabs*cflux*pow}).
Owing to the lack of statistical evidence for intrinsic absorption, a fixed Galactic neutral hydrogen column density of $N_{\rm H, \, MW}= 2.6 \times 10^{20} \, {\rm cm^{-2}}$
is adopted in the spectral analysis \cite{Willingale2013MN}\footnote{\url{https://www.swift.ac.uk/analysis/nhtot/index.php}}. 
The spectra from FXT-A and FXT-B are simultaneously fitted to improve statistical robustness and consistency.

The X-ray emission of AT 2024wpp was also observed by the X-ray Telescope (XRT) onboard on the {\it Neil Gehrels Swift Observatory}\cite{Gehrels2004ApJ, Burrows2005SSRv}, beginning on 2024 Sept.  27\cite{Srinivasaragavan2024TNS} (at $t\approx 2$ days post-discovery; PIs Coughlin and R. Margutti).
We processed the {\it Swift}-XRT data with HEASoft v.6.24, revealing a significant ($>3\sigma$) X-ray source coincident with the optical transient position during $\delta t \approx 2.1$--9.4 days.
The ancillary response file was generated with \texttt{xrtmkarf}, and the appropriate response matrix file (RMF) from the {\it Swift} CALDB was applied. The 0.5--10~keV spectrum was then extracted and grouped, following procedures analogous to those adopted for the {\it EP}-FXT data.

Additionally, we obtained four epochs of {\it Chandra X-ray Observatory (CXO)} ACIS-S observations of AT 2024wpp (PI R. Margutti) spanning $\delta t \approx +17.8$--75.1 days, which cover both the decay and rebrightening phases of the explosion\cite{Margutti2024TNS}.
The {\it CXO} data were reduced using CIAO v4.16 and the corresponding calibration files following standard procedures.
Consistent spectral analysis was performed by extracting 0.5--10 keV spectra and applying the same fitting methodology as for the {\it EP}-FXT data, thereby enabling direct comparison across all observations. The results of the spectral analysis for all X-ray observations are summarized in Table \ref{table:specx}.

\subsubsection*{X-ray analysis}~\label{sec:xray analysis}

\ref{fig:x ray light curve} presents the X-ray light curve of AT 2024wpp compared with those of other FBOTs.
Our observations reveal that the X-ray evolution of AT 2024wpp differs significantly from that of AT 2018cow and other known FBOTs.
The X-ray light curve exhibits three distinct phases: an initial rise, a subsequent decay, and a late-time rebrightening.
The X-ray luminosity initially rises between $\delta t \approx 2.1$ and $3.5$ days post-discovery, reaching a high peak luminosity of $L_{\rm X} \approx 2 \times 10^{43}$ ${\rm erg\, s^{-1}}$.
Following this peak, the luminosity enters a decay phase until $\delta t \approx 30$ days,
and is better described by a broken power-law model ($\chi^2=42.3$) than by a single power-law model ($\chi^2=72.0$).
Resembling the behavior observed in AT 2018cow\cite{Margutti2019}, the decay has a break time at $\delta t \approx 18$ days and consists of a slow decay ($\propto t^{-1.0}$) followed by a fast decay ($\propto t^{-4.9}$).
Finally, the light curve features a significant rebrightening starting at $\delta t \gtrsim 30$ days, reaching a secondary peak of $L_{\rm X} \approx 2 \times 10^{42}$ ${\rm erg\, s^{-1}}$.
This phenomenon, observed for the first time in an FBOT, suggest that a new outflow component emerges at late times.

The X-ray spectra of AT 2024wpp also exhibit peculiar evolutionary behavior. The temporal evolution of the spectral index $\beta$ (where $F_{\nu} \propto \nu^{-\beta}$) is displayed in \ref{fig:x ray light curve} and the X-ray spectra from three representative epochs are shown in \ref{fig:XraySpectrum}. 
Although the average spectral index of AT 2024wpp, $\left \langle \beta \right \rangle =0.66$, is similar to that of AT 2018cow\cite{Margutti2019}, its evolution is quite different.
Initially, the spectra are soft with $\beta>0$ at $\delta t \lesssim 13$ days (e.g., \ref{fig:XraySpectrum}a), and they then become hardened toward a nearly flat state ($\beta \approx 0$) during the fast decay phase (e.g., \ref{fig:XraySpectrum}b). 
Around the time of rebrightening peak ($\delta t = 49.6$ days), {\it CXO} captures an inverted spectrum with $\beta=-1.33$ (\ref{fig:XraySpectrum}c).
Following this peak, the spectra softened rapidly.
Furthermore, 
Ref.\cite{2025ApJ...993L...6N} reported a Compton hump in AT 2024wpp through combining data with hard X-ray observations from {\it NuSTAR}, which was observed in the initial stage of AT 2018cow\cite{Margutti2019} and possibly AT 2020mrf\cite{Yao2022ApJ}.
These distinct spectral behaviors further support the interpretation that a new emission component emerged at late times in AT 2024wpp.

\subsubsection*{Radio observations}~\label{sec:radio} Follow-up radio observations were also obtained for AT 2024wpp, with ATCA under project code CX582 (see Method 2.8). The array configuration of all observations had a maximum 6 km baseline (6A/6C). Our data were obtained at central frequencies of 5.5 GHz and 9.0 GHz, with each observation having a bandwidth of 2 GHz and lasting for 4 to 5 hr.

The Common Astronomy Software Application (CASA) was adopted for  standard data calibration. We created the individual images at central frequencies of 5.0 GHz and 6.0 GHz with a bandwidth of 1 GHz, and at 8.25 GHz, 8.75 GHz, 9.25 GHz, and 9.75 GHz with a bandwidth of 0.5 GHz, to obtain detailed spectral structure. The measured flux values are determined as the peak densities of the single Gaussian component fitted on the image regions by \texttt{CASA.imfit}. 
We started to analyse the radio data with a synchrotron self- absorption (SSA) model.
The radio SED can be fitted by a broken power law \cite{2002ApJ...568..820G, Yao2022ApJ},

\begin{equation}
    F_{\nu} = F_{\nu , \text{peak}} 
\left[ \left( \frac{\nu}{\nu_{\text{peak}}} \right)^{-s\beta_1} + \left( \frac{\nu}{\nu_{\text{peak}}} \right)^{-s\beta_2} \right]^{-1/s}\, ,
\end{equation}
where $F_{\nu , \text{peak}}$ is the peak flux, $\nu_{\text{peak}}$ is the peak frequency, $\beta_1$ is the optically thick index, $\beta_2$ is the optically thin index, and $s$ is the smoothing parameter. 


We approximate the synchrotron-emitting region as a sphere of radius $R$. For synchrotron self-absorption with a power-law electron distribution $N(E)\propto E^{-p}$ and an emission volume of $4\pi fR^3/3$ (where $f$ is the filling factor), the observed fluxes in the optically thick and thin regimes both scale with the source size $R$ and the magnetic field strength $B$ \cite{1998ApJ...499..810C}.
The peak frequency is the transition point between the optically thick and optically thin regimes of the spectrum. For the case that $p=3$ and $f=0.5$,
the radio-emitting region radius $R$, the magnetic field strength $B$, the post-shock energy density $U$, the CSM density $n_e$, and the mass-loss rate $\dot{M} (v_{\rm wind}=1000~{\rm km}/{\rm s})$, can be calculated as \cite{2019ApJ...871...73H, Yao2022ApJ}
\begin{equation}
R = 7.1 \times 10^{16} \left( \frac{\epsilon_e}{\epsilon_B} \right)^{-1/19} \left( \frac{L_{\nu,\text{peak}}}{(1+z)^4 10^{29} \text{ erg s}^{-1} \text{Hz}^{-1}} \right) ^{9/19}  \left( \frac{\nu_{\text{peak}}}{5 \text{ GHz}} \right)^{-1} \text{cm},
\end{equation}

\begin{equation}
B = 0.36\left( \frac{\epsilon_e}{\epsilon_B}\right)^{-4/19} \left( \frac{L_{\nu, {\rm peak}}}{(1+z)^410^{29} {\rm erg ~s^{-1}~Hz^{-1}}}\right)^{-2/19} \left( \frac{\nu}{5~ {\rm GHz}}\right) ~{\rm Gauss},
\end{equation}

\begin{equation}
U = 4.0\times 10^{48} \frac{1}{\epsilon_B}\left( \frac{\epsilon_e}{\epsilon_B}\right)^{-6/19} \left( \frac{L_{\nu, {\rm peak}}}{(1+z)^410^{29} {\rm erg ~s^{-1}~Hz^{-1}}}\right)^{23/19} \left( \frac{\nu}{5~ {\rm GHz}}\right)^{-1} ~{\rm erg},
\end{equation}

\begin{equation}
n_e = \left ( \frac{61}{\epsilon_B} \right) \left( \frac{\epsilon_e}{\epsilon_B} \right)^{-6/19} \left( \frac{L_{\nu, {\rm peak}}}{(1+z)^410^{29} {\rm erg ~s^{-1}~Hz^{-1}}}\right)^{-22/19}\left( \frac{\nu}{5~ {\rm GHz}}\right)^4  \left( \frac{\Delta t}{100~{\rm days}}\right)^2 {\rm cm}^{-3}.
\end{equation}

\begin{equation}
\dot{M} = 10^{-5} \frac{1}{\epsilon_B}  \left( \frac{\epsilon_e}{\epsilon_B} \right)^{-8/19} \left( \frac{L_{\nu, {\rm peak}}}{(1+z)^410^{29} {\rm erg ~s^{-1}~Hz^{-1}}}\right)^{-4/19}\left( \frac{\nu}{5~ {\rm GHz}}\right)^2  \left( \frac{\Delta t}{100~{\rm days}}\right)^2 {\rm M}_{\odot}~{\rm yr}^{-1} .
\end{equation}
\noindent
Here, $\epsilon_e$ and $\epsilon_B$ denote the fractions of shock energy deposited into relativistic electrons and magnetic fields, respectively \cite{2013MNRAS.436.1258H}. In our calculations, we adopt the equipartition assumption of $\epsilon_e = \epsilon_B = 1/3$ \cite{2022ApJ...932..116H}. We performed joints fits for the t~47 days and 85 days energy spectra by combining the radio data obtained at similar phases from Ref. \cite{Nayana+etal+2025+2024wpp_unprecedent_evolution}. Because of the limitation of the radio bandpass, during the fitting we fixed the parameter $s=1$. We assumed $\beta_2=-1$ when fitting the $\Delta t=47$ days energy spectrum, while fixing $\beta_1=2.5$ and $\beta_2=-1$ for the $\Delta t=108$ days spectrum, as are typical of an SN in the optically thin regime \cite{2022ApJ...926..112B, 2024A&A...691A.329C}. The data points corresponding to $\Delta t$ = 137 days were excluded from the analysis because of their poor signal-to-noise ratios. 
The fits were performed using the 
MCMC sampler \texttt{EMCEE} \cite{2013PASP..125..306F}, with the best-fit results reported in \ref{tab1}. The derived averaged shock velocity is $v_{\rm sh}\gtrsim 0.2c$.

\subsubsection*{Model analysis and discussion}
Several models have been proposed to explain the physical origin of 18cow-like FBOTs, including explosion ejecta interactions with CSM \cite{Leung}, a BH tidally disrupting a star, 
the merger of a compact object
and massive star, CCSN, accreting compact object, and magnetar-powered models. However, none of these models can fully account for all observational features. While CSM interaction effectively explains the rapid rise and decline of the optical and radio light curves, it fails to reproduce the 
quasi-periodic
oscillations in the soft X-ray band. Additionally, the extreme luminosity and high velocities challenge standard SN models. Currently, the most plausible explanation involves a compact central engine --- either a magnetar, or an accreting neutron star or BH.

\subsubsection*{Classical power models}
\label{sssec: models}
Considering the energy diffusion into homologously expanding SN ejecta (mass $M_{\rm ej}$, velocity $v_{\rm ej}$, and optical opacity $\kappa$) \cite{1982ApJ...253..785A}, the SN luminosity powered by radioactive decay of $^{56}$Ni and $^{56}$Co can be written as 
\begin{equation}
L_\mathrm{rd}(t) = e^{-\left( t/t_{\rm diff} \right)^2}\int_0^t 2 M_{\rm{Ni}} 
 \left[ (\epsilon_{\rm{Ni}} - \epsilon_{\rm{Co}}) e^{-t' / t_{\rm{Ni}}} + \epsilon_{\rm{Co}} e^{-t' / t_{\rm{Co}}}\right] \left(1-e^{-A/t'^{2} }\right) e^{\left( t'/t_{\rm diff} \right)^2} \frac{t'dt'}{t_{\rm diff}^2},
\end{equation}
where $t$ is the time relative to explosion date, $t_{\rm diff}=\left(2 \kappa M_{\rm ej}/\beta c v_{\rm ej}\right)^{1/2}$ is the diffusion time with $\beta=13.8$ and $c$ being the light speed, and $A=3 \kappa_\gamma M_{\rm ej}/4 \pi v_{\rm ej}^2$ is the leaking factor with $\kappa_\gamma$ being the opacity for the photons from the energy source \cite{2015ApJ...799..107W, 2017ApJ...850...55N}. Here $M_{\rm{Ni}}$ is the mass of newly synthesized nickel, $t_{\rm{Ni}}$ and $t_{\rm{Co}}$ are the respective decay times of $^{56}$Ni and $^{56}$Co, and $\epsilon_{\rm{Ni}}$ and $\epsilon_{\rm{Co}}$ are the respective energy-generation rates \cite{Valenti2008MNRAS.383.1485V}. 

In the magnetar-powered scenario, the SN luminosity can be expressed as 
\begin{equation}
L_{\rm m}(t) = e^{-\left( t/t_{\rm diff} \right)^2}\int_0^t 2 \, F_{\rm mag} \left(1-e^{-A/t'^{2} }\right) e^{\left( t'/t_{\rm diff} \right)^2} \frac{t'dt'}{t_{\rm diff}^2},
\end{equation}
where $F_{\rm mag}$ denotes the input wind power from a central magnetar with an initial period $P_0$ and magnetic field strength $B$ \cite{2010ApJ...717..245K}. 

The formula for the SN luminosity powered by fallback accretion is given by
\begin{equation}
L_{\rm fb}(t) = e^{-\left( t/t_{\rm diff} \right)^2}\int_0^t 2 \, F_{\rm acc}\left(1-e^{-3 \kappa M_{\rm ej}/{4 \pi v_{\rm ej}^2t'^{2}} }\right)  e^{\left( t'/t_{\rm diff} \right)^2} \frac{t'dt'}{t_{\rm diff}^2},
\end{equation}
where $F_{\rm acc}=L_\mathrm{0,acc}\left[1+(t-t_\mathrm{i,acc})/ t_\mathrm{d,acc}\right]^{-5/3}$ is the input power from fallback accretion. 

The SN luminosity powered by ejecta-CSM interaction and radioactive decay of $^{56}$Ni+$^{56}$Co is \cite{2012ApJ...746..121C}
\begin{equation}
L_{\rm csird}(t) = e^{-t/t_0}\int_0^t \left(F_{\rm fs}+F_{\rm rs}\right)  e^{ t'/t_0} \frac{dt'}{t_0}+e^{-t/t_{0,1}}\int_0^t F_{\rm rd}e^{ t'/t_0} \frac{dt'}{t_{0,1}},
\end{equation}
where $F_{\rm fs}$ and $F_{\rm rs}$ are the input powers from forward and reverse shocks produced in the ejecta-CSM interaction, $F_{\rm rd}$ is the input power from radioactive decay of $^{56}$Ni and $^{56}$Co, and $t_0$ and $t_{0,1}$ are the respective diffusion timescales for interaction and decay energy.

The MCMC algorithm is performed to obtain best-fit parameters for the above models. The fitting results are shown in \ref{fig: model_sim}, with the corresponding parameters listed in \ref{tab: model_sim}.

\subsubsection*{Accreting magnetar model}

To interpret the luminosity evolution of the thermal optical and soft X-ray emission of AT 2024wpp, we adopt an accreting magnetar engine model that builds upon the semianalytic framework developed in earlier work \cite{2024ApJ...963L..13L}. This model considers a two-zone configuration: an inner stripped magnetar wind is subject to gradual magnetic dissipation, and an outer expanding ejecta is responsible for thermalization and reprocessing. 
In this framework, the high-energy radiation arises from gradual magnetic dissipation within the relativistic stripped wind launched by the newborn magnetar, which produces the observed soft X-ray emission \cite{2017MNRAS.468.3202B,2017ApJ...846..130X}. Meanwhile, a fraction of the dissipated energy is deposited into the surrounding ejecta, where it thermalizes and diffuses outward as optical–UV radiation \cite{2024ApJ...963L..13L}. The optical light curve thus traces the reprocessed emission powered by the magnetar’s spin-down energy, while the soft X-rays directly probe the dissipation in the inner magnetar wind region.

We further develop this model by including the contribution of late-time fallback accretion onto the newborn magnetar. The accretion onto the magnetar is modeled following theoretical prescriptions \cite{2011ApJ...736..108P,2012ApJ...759...58D,2021ApJ...914L...2L}. The accretion rate is parameterized as
\begin{equation}
    \dot{M}(t) = \dot{M}_0 \left( \frac{t}{t_0} \right)^{-\alpha}\, ,
\end{equation}
where $\alpha=5/3$ is a typical power-law index for fallback-dominated accretion. In our model, fallback accretion transfers angular momentum to the newborn magnetar and spins it up, thereby enhancing the magnetic dipole spin-down luminosity. This increased spin-down power produces the significant late-time rebrightening observed in the optical and soft X-ray light curves.

To constrain model parameters, we perform 
MCMC fitting using the observed optical bolometric light curve and the 0.5–10 keV X-ray luminosity. The best-fit solution indicates an ejecta mass of $\sim 0.66~M_\odot$ and a velocity of $\sim 0.14c$. The central engine is a newborn millisecond magnetar with an initial spin period $P_0 \approx 4.6~\rm ms$ and surface dipole magnetic field $B_p \approx 6\times10^{14}~\rm G$. The onset time of fallback accretion is constrained as $t_0 \approx 41~{\rm days}$, with an initial accretion rate $\dot{M}_0 \approx 1.4 \times 10^{-6}~ {\rm M}_\odot~ {\rm s}^{-1}$.
The accreting magnetar model is able to reproduce the overall morphology of both the bolometric and soft X-ray luminosity evolution (Extended Data Fig.~10). Systematic residuals remain: in the bolometric light curve the model is higher than the observed values at $\sim 40$--70 days, and it slightly overestimates the X-ray luminosity at $\sim 65$ days.
Overall, the results suggest that the accreting magnetar model provides a reasonable description of the observed luminosity evolution of AT 2024wpp, provided that a late-onset fallback accretion component is present.

\subsubsection*{The delayed merger of a massive star and a black hole}

The delayed merger model for FBOTs was first proposed in Ref. \cite{Metzger2022ApJ}. 
Nevertheless, this model faces several challenges when applied to AT 2024wpp: it fails to reproduce the detailed X-ray luminosity evolution, the optical light curve is poorly fitted if the emission is entirely due to X-ray reprocessing, and the rapid X-ray fluctuations are more naturally attributed to direct radiation from the central engine rather than in reprocessing by optically thick ejecta.
To overcome these limitations, we modify this model for AT 2024wpp to be able to account for its bolometric and X-ray light curves, as illustrated in \ref{fig:Model}.
In this scenario, a massive star, such as a WR star, merges with a BH companion hundreds to thousands of years after the initial common-envelope (CE) phase, eventually leading to a violent explosion.
The premerger phase begins when the CE event strips the WR star of its H-rich envelope, forming a distant, H-rich CSM.
Subsequently, a remnant disk extracts angular momentum from the binary, which drives the exposed He core of the WR star to be tidally disrupted by the BH.
This process initiates Roche-lobe overflow (RLOF), resulting in unstable mass transfer and thus forming He-rich CSM surrounding the binary.

The disrupted star is accreted onto a rapidly spinning BH, generating a strong magnetic field that powers a sustained outflow through the Blandford-Znajek (BZ) mechanism\cite{Blandford1977MNRAS} by extracting the BH spin energy.
This outflow is anisotropic, with an inclination angle-dependent velocity and density that features a faster component along the polar regions, similar to the scenario in TDEs\cite{Dai2018ApJ}. 
This geometry provides the foundation for a two-component kinematic model that successfully reproduces the spectral evolution\cite{Aspegren2026arXiv}.
The initial optical light is primarily generated by the shock interaction between the low-mass polar ejecta and the CSM, supplemented by the reprocessing of X-rays to maintain a persistently high temperature of LFBOT emission over several weeks\cite{Klencki2025arXiv}. 
The evolution of the radiation energy ($E_{\rm ej}$) from the polar ejecta can be described by the energy conservation equation,
\begin{equation}\label{opt1} 
	\frac{\mathrm{d}  E_{\mathrm{ej}}}{\mathrm{d} t}
	= -\frac{E_{\mathrm{ej}}}{t}-L_{\mathrm{bolo, ej}} + L_{\mathrm{sh,ej}}  +
	L_{\mathrm{acc}}\, ,
\end{equation}
where the terms on the right-hand side account for adiabatic expansion losses ($E_{\mathrm{ej}}/t$), radiative losses ($L_{\mathrm{bolo, ej}}$), shock heating ($L_{\mathrm{sh,ej}}$), and accretion power ($L_{\mathrm{acc}}$), respectively.
The radiative loss term is given by
\begin{equation}
	L_{\mathrm{bolo, ej}}
	= \frac{E_{\mathrm{ej}}}{t_\mathrm{diff, ej} + t_\mathrm{lc, ej}}\, ,
\end{equation}
where $t_\mathrm{diff, ej}
= 3 M_\mathrm{ej} \kappa_\mathrm{ej} / (4 \pi R_\mathrm{ej} c) $ is the photon diffusion time and $t_\mathrm{lc, ej} = R_\mathrm{ej}/c = v_\mathrm{ej} t/c$ is the light-crossing time. 
Here, $M_\mathrm{ej}$ and $v_\mathrm{ej} $ are respectively the mass and velocity of the polar ejecta, respectively,
and we assume a constant opacity of $\kappa_\mathrm{ej}=0.2\; \mathrm{cm^2 \; g^{-1}}$.
The luminosity injected by the shock heating term is given by
\begin{equation}
	L_{\mathrm{sh}} = 4 \pi R_{\mathrm{sh}}^{2} \rho_{\mathrm{sh}}
	\left(v_{\mathrm{sh}}^{2} / 2\right) v_{\mathrm{sh}}\, ,
\end{equation}
where $R_{\mathrm{sh}} \approx R_\mathrm{ej} $ is the shock radius. 

In the initial stage of the explosion, when the ejecta mass is greater than that of the swept-up CSM, the deceleration of the ejecta can be neglected; we thus assume a constant shock velocity $v_{\mathrm{sh}} \approx v_\mathrm{ej}$.
The density profile of the premerger CSM, $\rho_{\mathrm{sh}}$, is modeled as\cite{Metzger2022ApJ,  Metzger2017MN}
\begin{equation} 
	\rho_{\mathrm{sh}} = \frac{\dot{M}_{\mathrm{esc}}}{4 \pi v_\mathrm{esc} r^2} 
	\exp \left[-\frac{r}{v_\mathrm{esc} t_\mathrm{esc} } \right]\,  .
\end{equation}
Here $\dot{M}_{\mathrm{esc}}$ is the mass-loss rate of the escaping H-poor matter before merger;
$v_{\mathrm{esc}}$ and $t_\mathrm{esc}$ are the escape velocity and time, respectively. 
We set the escape velocity to $v_{\mathrm{esc}}  = 0.02 c$.
For the last term, the energy injected from hyperaccretion is given by
\begin{equation} 
	L_{\rm acc} =\eta_{\rm acc} \dot{M}_{\rm acc}  c^2\, ,
\end{equation}
where $\eta_{\rm acc}$ is the efficiency of converting accretion energy to thermal radiation. 
The accretion rate $\dot{M}_{\rm acc}$ is parametrized as a power-law function of radius and time\cite{Metzger2022ApJ, Metzger2008MN}, 
\begin{equation}\label{Macc} 
	\dot{M}_{\rm acc} = \frac{\dot{M}_0}{3} \left( \frac{R_\mathrm{in}}{R_\mathrm{d}} \right) ^p
	\left( \frac{ t }{ t_\mathrm{visc} } \right)^{-\frac{4(p+1)}{3}} , \, t>t_\mathrm{visc} ,
\end{equation}
where $R_\mathrm{d} = a_{\mathrm{RLOF}}/(1+M_{\mathrm{star}}/M_\mathrm{BH})$ is the characteristic radius of the disk\cite{Margalit2016MN}, $a_{\mathrm{RLOF}}$ is the RLOF radius\cite{Eggleton1983ApJ}, and $M_{\mathrm{star}}$ and $M_\mathrm{BH}$ respectively denote the mass of the star and BH.
Also, $R_\mathrm{in} = 6 G M_{\rm BH} / c^2$ is the inner edge of the disk,  represented by the innermost stable circular orbit of the Schwarzschild BH. 
For the index $p$, we adopt a value of 0.6 in this work, motivated by numerical simulations of radiatively inefficient accretion flows\cite{Blandford1999MN, Yuan2014ARAA, Hu2022ApJ}. 
 The peak accretion rate near the outer disk $R_\mathrm{d}$ is $\dot{M}_0 = M_\mathrm{d} /t_\mathrm{visc} $, assuming a disk mass of $M_\mathrm{d} \approx M_\mathrm{star}$.
$t_{\mathrm{visc}}$ is the viscous timescale, given by $t_{\mathrm{visc}}\approx  \alpha^{-1} \theta^{-2} 
(R_{\mathrm{d}}^{3} / G M_\mathrm{BH})^{1 / 2} $. 
Here we set $\alpha=0.1$ as the viscosity parameter and $\theta = H/R_{\rm d} = 1$ as the aspect ratio for the thick disk\cite{Metzger2022ApJ, Tuna2023ApJ}.

With a low-velocity outflow (or wind) present in the nonpolar direction, similarly to the first optical component (Eq.~\ref{opt1}), the evolution of the radiation energy ($E_{\mathrm{w}}$) can be described by
\begin{equation}\label{opt2}
	\frac{{\rm d} E_{\mathrm{w}}}{{\rm d} t}
	= -\frac{E_{\mathrm{w}}}{t}-L_{\mathrm{bolo, w}}  + L_{\mathrm{sh,w}}\, .
\end{equation}
The terms on the right-hand side of the equation represent adiabatic expansion losses ($E_{\mathrm{w}}/t$), radiative losses ($L_{\mathrm{bolo, w}}$), and shock heating ($L_{\mathrm{sh,w}}$), respectively.
The bolometric radiative loss is defined as 
\begin{equation} 
	L_{\mathrm{bolo, w}}
	= \frac{E_{\mathrm{w}}}{t_\mathrm{diff, w}}\, ,
\end{equation}
where $t_\mathrm{diff, w}
= 3 M_\mathrm{w} \kappa_\mathrm{w} / (4 \pi R_\mathrm{w} c) $ is the photon diffusion time;
$M_\mathrm{w}$ and $v_\mathrm{w}$ are the mass and velocity of the slow outflow, respectively. 

An accreting spinning BH can maintain a strong, large-scale poloidal configuration magnetic field.
This configuration can drive a mildly relativistic outflow via the BZ mechanism, which is the source of the bright X-ray emission. 
A characteristic timescale for the BZ jet activity is determined by the balance between two pressures at the magnetospheric radius. 
The magnetic pressure and the ram pressure of the accretion flow are denoted as 
\begin{equation}\label{PB}
	P_B = \frac{B_\mathrm{H}^2}{8 \pi} \left( \frac{r}{r_H} \right)^{-4},
\end{equation}
and
\begin{equation}\label{Pf}
	P_f = \frac{G M_\mathrm{BH} \dot{M}}{2 \pi r^3 v_r} ,
\end{equation}
respectively. Here $B_\mathrm{H}$ is the magnetic field strength at the BH horizon, $r_H = 2 G M_\mathrm{BH} /c^2$ is the radius of the outer horizon, and $v_r = \epsilon (G M_\mathrm{BH}/r)^{1/2}$ is the inflow radial velocity, where the dimensionless parameter $\epsilon $ is taken to be $\sim 0.01$ as supported by observations and numerical simulations\cite{Tchekhovskoy2011MN, Zamaninasab2014Natur}. 
As the accretion rate decreases over time, the ram pressure of the matter $P_f$ drops below the magnetic pressure $P_B$. 
The point at which these pressures are equal ($P_f=P_B$) defines a break time in the X-ray light curve.
Combining Eq.~(\ref{Macc}), Eq.~(\ref{PB}), and Eq.~(\ref{Pf}), we can obtain the characteristic break timescale,
\begin{eqnarray}\label{tbreak}
	t_{b} &=&
	\left( \frac{4 c^5}{3 \epsilon G^2} \right)^{ \hat{\alpha}}	
	\left(\frac{M_\mathrm{star} }
	{ M_\mathrm{BH}^{2} B_{H,0}^{2}	}\right)^{ \hat{\alpha}}
	\left(\frac{R_\mathrm{in}}{R_\mathrm{d}}\right)^{\hat{\alpha}p}
		t_\mathrm{visc}^{1-\hat{\alpha}}  \nonumber \\
		&\approx & 16 \, {\rm days} \left( \frac{M_{\rm star}}{30 \, M_{\odot}} \right)^{ \hat{\alpha}} 
		\left( \frac{M_{\rm BH}}{10 \, M_{\odot}} \right)^{ -2 \hat{\alpha}} 
		\left( \frac{B_{\rm H,0}}{10^{12} \, \rm G} \right)^{ -2 \hat{\alpha}}
		\left( \frac{R_\mathrm{in}/ R_\mathrm{d}}{10^{-4} } \right)^{ \hat{\alpha}p} 
		\left( \frac{t_\mathrm{visc}}{0.1 \, \rm day} \right)^{ 1-\hat{\alpha} }\, , 
\end{eqnarray}
where we define an index $\hat{\alpha} \equiv 3/(4p+4) \overset{ {}_{p=0.6}}{=} 15/32 \approx 1/2$.
The typical values for the disk geometry ($R_\mathrm{in}/ R_\mathrm{d}\approx 10^{-4}$) and viscous timescale ($t_\mathrm{visc}\approx 0.1\, {\rm day} $) refer to Ref.\cite{Metzger2022ApJ}.
The break time of $16 \, {\rm days}$ aligns well with the steep decay phase in the X-ray observations. 
It marks the epoch when the declining accretion rate reduces the ram pressure of the inflow, allowing the magnetosphere to push outward until magnetic and ram pressures balance at the magnetospheric radius $R_m$. 
This transition leads to a magnetically arrested disk (MAD)\cite{Igumenshchev2003ApJ, Igumenshchev2008ApJ, Tchekhovskoy2015MNRAS, You2023Sci}, accompanied by an outward expansion of $R_m$ and a rapid decline of the BH-threading magnetic flux $\Phi_B$, ultimately shutting off the BZ-powered jet\cite{Wu2013ApJ, Tchekhovskoy2015MNRAS, Kisaka2015ApJ, Zheng2024ApJ}.
Therefore, the maximum power of the BZ mechanism evolves as
$L_{\mathrm{BZ}} \approx \Phi_B^2 \approx R_{m}^{-4} \approx t^{-16/3}.$
Within this scenario, the X-ray evolution can be parameterized to a smoothly broken power-law function, 
\begin{equation}
	L_{\mathrm{X, BZ–1}}  =  e^{-\tau_{\rm X, 1}}  L_0
	\left[ \left( \frac{t}{t_{b} } \right)^{sk} + \left( \frac{t}{t_{b} } \right)^{s16/3} \right]^{-1/s}\, ,
\end{equation}
where $\tau_{\rm X, 1}=M_{\rm ej}^{\prime} \kappa_{\rm ej}/ (4 \pi R_{\rm  ej }^2 )$ is the optical depth for X-ray emission and $s=2$ is the smoothing parameter. 
Given that the X-ray luminosity peaks earlier than the optical, with timescale $t_{\rm pk,X}\approx t_{\rm pk,opt}/4$, which implies a lower column density along the X-ray escape direction, we thus adopt $M_{\rm ej}^{\prime}=M_{\rm ej}/4$.
Above, $k$ is the decay slope before the break time, which connects the evolution of the BH spin and the radius of the outer horizon. The characteristic X-ray luminosity at $t_{b}$ is given by
\begin{eqnarray}
	L_{\mathrm{0}} & = & \eta_{\rm BZ}  \frac{\pi c}{320}  r_H^2 B_{H,0}^2  \nonumber \\
	& \approx & 6 \times 10^{42} \, \mathrm{erg\, s^{-1}}
	\left( \frac{\eta_{\rm BZ}}{10^{-2}} \right)
	\left( \frac{M_{\rm BH}}{15 \, M_{\odot}} \right)^2
	\left( \frac{B_{\rm H,0}}{10^{12}\, \rm G} \right)^2\, .
\end{eqnarray}
This luminosity is comparable to the values observed in the X-ray light curve.

At late times, a portion of the pre-explosion equatorial outflow from the L2 point remains gravitationally bound and falls back toward the BH, leading to an increased accretion rate (\ref{fig:Model}(d)). 
The energy released from this fallback accretion process fuels a secondary BZ jet, which drives the observed X-ray rebrightening\cite{Wu2013ApJ} and the slight bump in the optical light curve. 
The fallback accretion rate is described by a single power law,
\begin{equation}\label{Mfb}
	\dot{M}_{\rm fb}(t)= \dot{M}_{\rm fb}  \left(\frac{t}{ t_{\rm fb}} \right)^{-5 / 3} , \, t>t_{\rm fb}\, ,
\end{equation}
where $M_{\mathrm{fb}}$ is the total fallback mass and $t_{\rm fb}$ is the time at which fallback accretion begins.
This timescale can be estimated by $t_{\rm fb} = \int dr/v_r \approx R_{\rm fb} / v_r$, which gives 
\begin{equation}\label{Rfb}
	t_{\rm fb}\approx  32 \, {\rm days} \left( \frac{\epsilon}{10^{-2}} \right)^{-1} \left(\frac{R_{\rm fb} }{10^{12} \, {\rm cm} } \right)^{3 / 2}
	\left(\frac{{M}_{\rm BH} }{ 10 \, {\rm M}_{\odot}} \right)^{-1 / 2} ,
\end{equation}
where $R_{\rm fb}$ is the initial distance of the fallback matter,  comparable to the location of the outer Lagrangian point L2 at radius $R_{\rm L2} \approx a_{ \rm RLOF} $\cite{Lu2023MN, Schneider2025arXiv} during the premerger phase. 
L2 is an unstable gravitational equilibrium point, which allows matter to accumulate and outflow\cite{Lu2023MN, Schneider2025arXiv}.
According to the simulation, in binaries with mass ratio $q\gtrsim 0.8$ and cold mass loss, the equatorial outflow from L2 remains marginally bound and can fall back onto the binary system over a timescale of tens to hundreds of orbital periods \cite{Pejcha2016MNRAS}.
Thus, it is plausible that a portion of this equatorial outflow falls back toward the central object, contributing to the rebrightening observed in AT 2024wpp. 
Note that this timescale is sensitive to the parameter $\epsilon$. 
According to accretion theory, the radial velocity $v_r$ is less than the sound speed $c_s$: $v_r \ll (H/R_d) c_s \approx 10\, {\rm km \, s^{-1}} $\cite{Yuan2014ARAA}.
Therefore, the value $\epsilon=0.01$ used here is a reasonable assumption\cite{Yuan2014ARAA, Kisaka2015ApJ}.

During the fallback phase, the inner disk enters a MAD state.
By combining the pressure balance $P_B=P_f$ (Eq.~\ref{PB} and Eq.~\ref{Pf}), we obtain the magnetic field intensity in the MAD as 
\begin{equation}\label{BMAD}
	B_{\rm MAD} = 2 G^{1/4} \epsilon^{-1/2} M_{\rm BH}^{1/4} \dot{M}_{\rm fb}(t)^{1/2} R_m(t)^{-5/4}\, .
\end{equation}
The light curve for the X-ray rebrightening component is modeled similarly to the first one: $L_{\rm X, BZ–2}=\eta_{\rm BZ}  e^{-\tau_{\rm X, 2} }  L_{\rm BZ} \propto B_{\rm MAD}^2$, where $\tau_{\rm X, 2}=M_{\rm w} \kappa_{\rm w}/ (4 \pi R_{\rm  w }^2 )$ is the X-ray optical depth. 
Solving Eq.~\ref{opt1} and Eq.~\ref{opt2}, the bolometric luminosity for the ejecta ($L_{\rm  bolo, ej}$) and the slow outflow ($L_{\rm bolo, w}$) can be obtained.
Given that the optical bump is comparable in luminosity to the second X-ray peak, we assume $L_{\rm  bolo, fb} \approx L_{\rm X, BZ–2}$. Eventually, the total bolometric luminosity and X-ray luminosity are then calculated as the sum of their components,
\begin{equation} 
	\begin{cases}
		L_{\rm bolo} = L_{\rm bolo, ej} + L_{\rm bolo, w} + L_{\rm bolo, fb},
		\\
		L_{\rm X} = L_{\rm X, BZ–1} + L_{\rm X, BZ–2}.
	\end{cases}.
\end{equation}
We employ an MCMC analysis to fit this model using the observed optical bolometric and the 0.5–10 keV X-ray luminosity data. The free parameters, priors and resulting constraints on the WR/BH merger model are listed in Table \ref{tab:fit_results} and the best-fit light curves are displayed in \ref{fig:Model}(a).
The constraint results reveal that the overall optical and X-ray luminosity evolution of  AT 2024wpp can be explained by the scenario involving the merger of a $\sim 15\,{\rm M}_{\odot}$ BH and a WR star with an initial mass $\sim 34\, {\rm M}_{\odot}$. 
The merger-driven explosion generates low-mass ejecta of $\sim 0.3\, {\rm M}_{\odot}$ with a mildly relativistic velocity of $\sim 0.2c$.
Subsequently, the fallback accretion of $0.1 \, {\rm M}_{\odot} {\rm yr}^{-1}$ after $\gtrsim 30$ days causes the  rebrightening in both X-ray and optical/UV bands.







\begin{figure}
    \centering
    \includegraphics[width=0.7\textwidth]{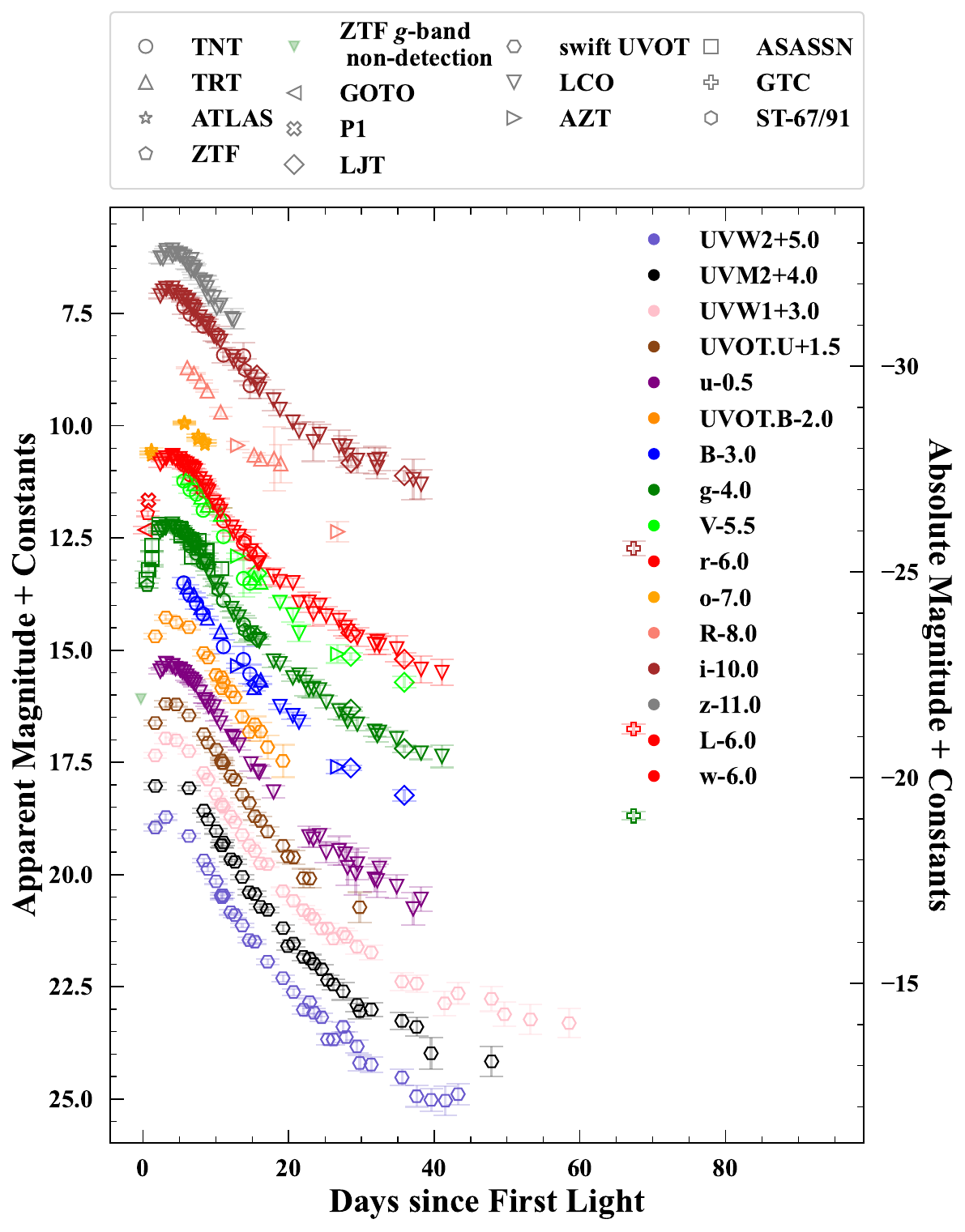}
    \caption{\textbf{UV and optical light curves of AT\,2024wpp until $\sim+$70 days relative to the estimated time of first light.} 
    Different symbols represent various facilities that were used to obtain the photometry (see top legend). Data in different filters are shown with different colors and shapes, and are shifted vertically for clarity, as indicated in the left legend. The data point with a downward arrow represents the nondetection limit of the ZTF $g$ band \cite{2024TNSTR3719....1S}, obtained 0.623 days before the earliest detection by GOTO $L$ and 0.935 days by ZTF $g$ (\ref{table:all_lc}); we assumed the first-light time to be the average of these two epochs,  $t_0=60,578.75\pm0.31$ (Sep. 25.75). Note that, in contrast to Ref.\cite{2024TNSTR3719....1S}, Ref.\cite{Perley2026arXiv} reported a low S/N detection of the ZTF $g$ band on MJD 60,578.437, which is close to the $t_0$ adopted in our analysis. Our conclusions are robust against variations in the adopted explosion time.
    }
    \label{fig:all_lc}
\end{figure}

\begin{figure}[htbp]
\centering
\begin{minipage}[b]{0.5\textwidth}
    \centering
        \begin{overpic}[width=0.98\textwidth]{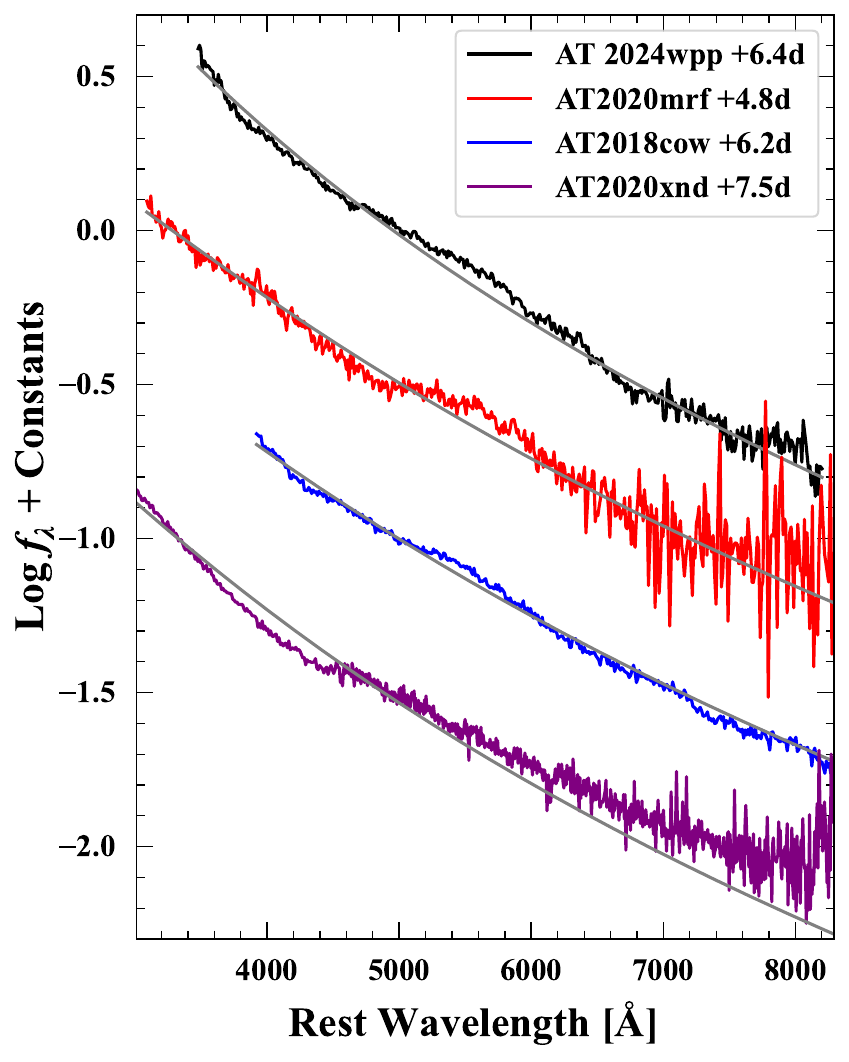}
        \put(0,90){\textbf{a}}
        \end{overpic}
\end{minipage}
\begin{minipage}[b]{0.48\textwidth}
    \centering
        \begin{overpic}[width=0.98\textwidth]{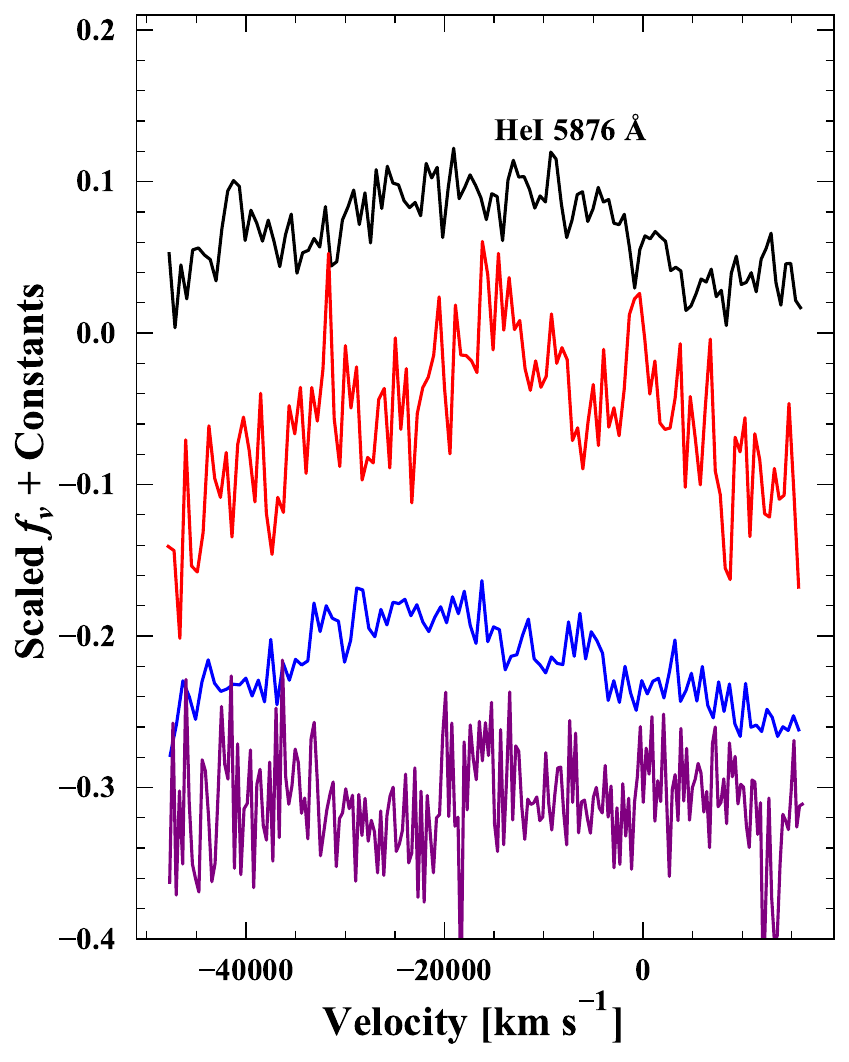}
        \put(0,90){\textbf{b}}
        \end{overpic}
\end{minipage}
\caption{
\noindent\textbf{Comparison of the bump in the 5000--6000 \AA\ region between AT~2018cow\cite{2021ApJ...910...42X}, 2020mrf\cite{2020TNSCR1846....1B}, 2020xnd\cite{2021MNRAS.508.5138P}, and 2024wpp.} The spectra are rebinned to 10 \AA\ to increase the S/N.
(a) Blackbody fits to the spectra of AT~2018cow, 2020mrf, 2020xnd, and 2024wpp at $t \approx +6$ days after  first light. All spectra have been corrected for reddening and host-galaxy redshift. Compared with the blackbody continuum, a broad bump is visible in the 5000--6000 \AA\ region for AT~2018cow, 2020mrf, and 2024wpp. No clear bump can be seen in the spectrum of AT~2020xnd, though the blackbody fit is not good. 
(b) The 5000--6000 \AA\ region, after subtraction of the blackbody continuum shown in the left panel, in velocity space relative to the rest wavelength of He\,{\sc i} 5876 \AA. 
}
\label{fig:He5876}
\end{figure}

\begin{figure}
    \centering
    \includegraphics[width=0.73\textwidth]{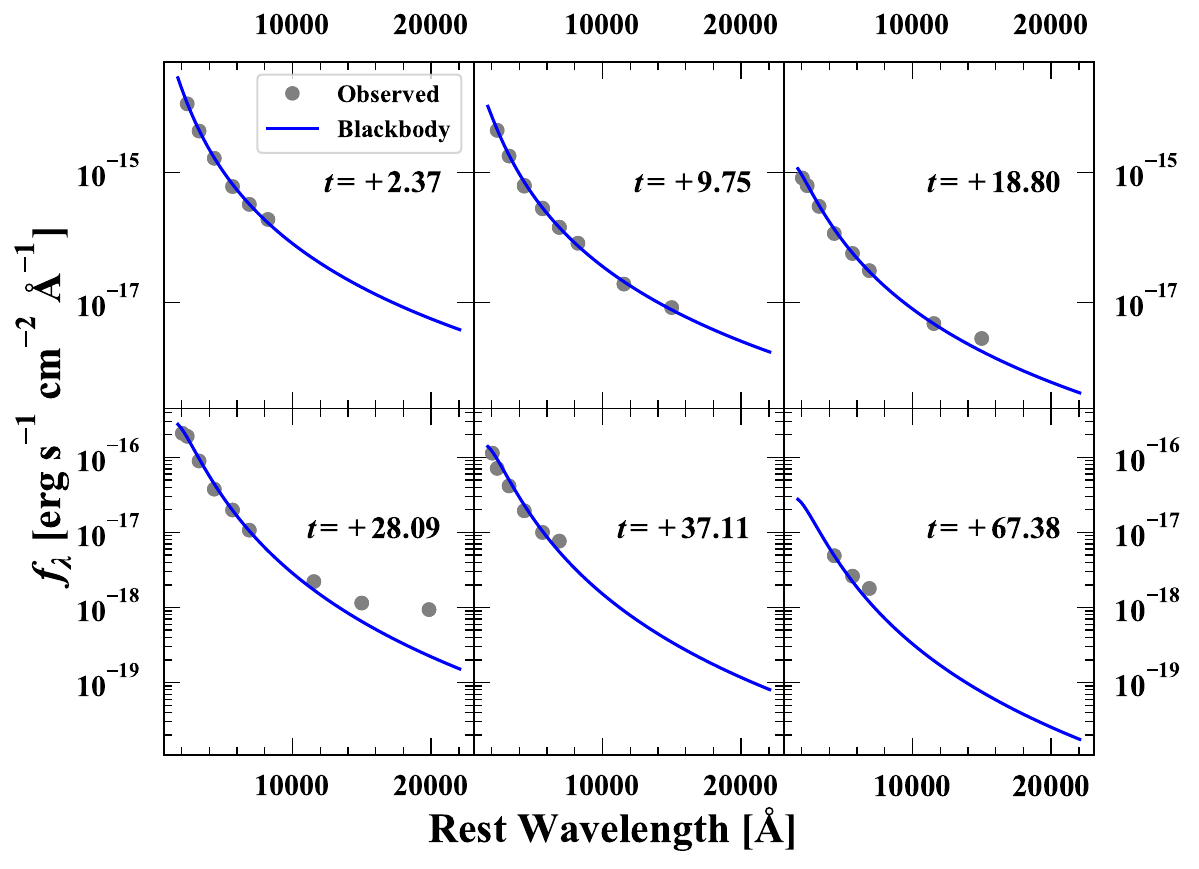}
    \caption{\textbf{SED evolution of AT~2024wpp.} Photometric observations are shown as grey circles and the blackbodies estimated in Section\,\ref{sec:Bol_Temp} are shown as blue curves. The flux density has been corrected for host-galaxy redshift and Galactic reddening.  
    }
    \label{fig:SED_fit}
\end{figure}

\begin{figure}
    \centering
    \includegraphics[width=0.8\columnwidth]{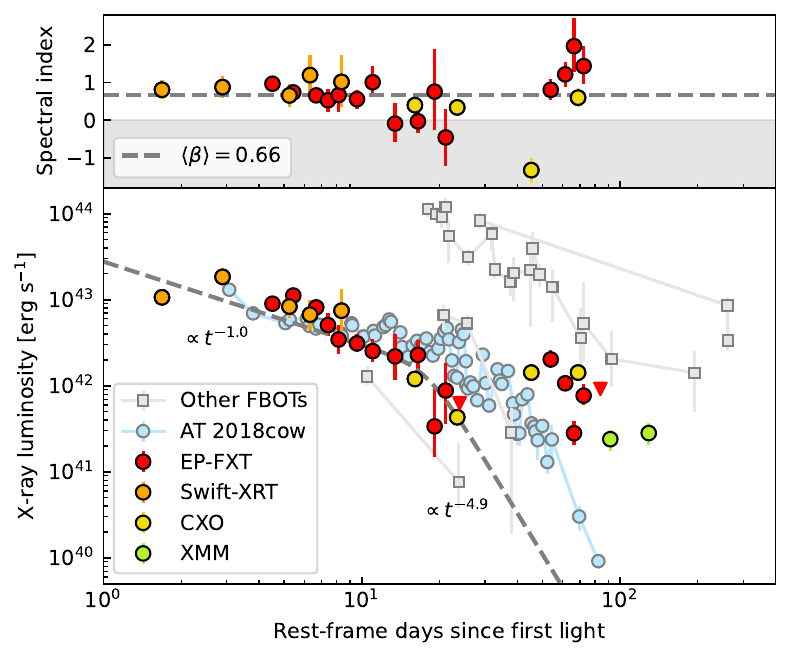}
    \caption{\textbf{The X-ray spectral index evolution and light curve of AT 2024wpp.}
	{\it Upper panel:} Temporal evolution of the spectral index $\beta$ (where $F_{\nu} \propto \nu^{-\beta}$).
	The horizontal dotted line indicates the average spectral index $\left \langle \beta \right \rangle =0.66$.
	The grey shaded region highlights epochs where $\beta<0$, corresponding to inverted spectral indices.
	{\it Lower panel:} The X-ray luminosity light curve in the 0.5--10 keV band for AT 2024wpp, including data from {\it EP}-FXT (red dots), {\it Swift}-XRT (orange dots), {\it CXO} (yellow dots) and {\it XMM-Newton} (green dots, from Ref. \cite{2025ApJ...993L...6N}).
	The decay phase at $\delta t \lesssim 30$ days can be well described by a broken power-law model with decay slopes of $t^{-1.0}$ and $t^{-4.9}$ (grey dotted line).
	For comparison, the light curve of the archetypal FBOT AT 2018cow is shown\cite{Margutti2019}, along with other FBOTs including AT 2020xnd\cite{Bright2022ApJ}, AT 2020mrf\cite{Yao2022ApJ}, AT 2022tsd\cite{Ho2023Nature}, and AT 2023fhn\cite{Chrimes2024AA} (grey squares, not  distinguished).}
    \label{fig:x ray light curve}
\end{figure}

\begin{figure}[htbp]
\centering
\begin{minipage}[b]{0.325\textwidth}
    \centering
        \begin{overpic}[width=1\textwidth]{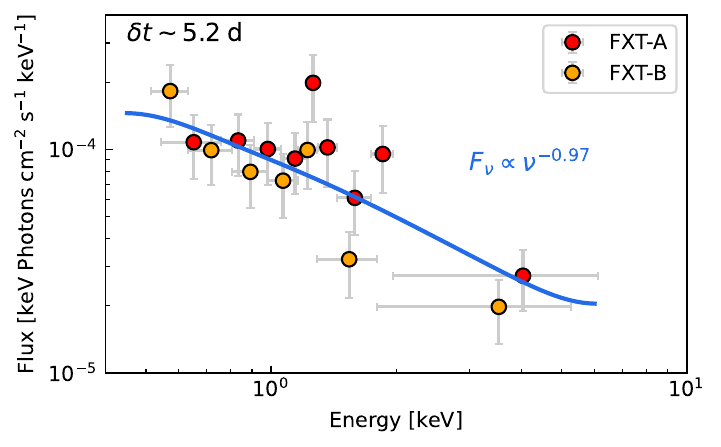}
        \put(-2,61){\textbf{a}}
        \end{overpic}
\end{minipage}
\begin{minipage}[b]{0.325\textwidth}
    \centering
        \begin{overpic}[width=1\textwidth]{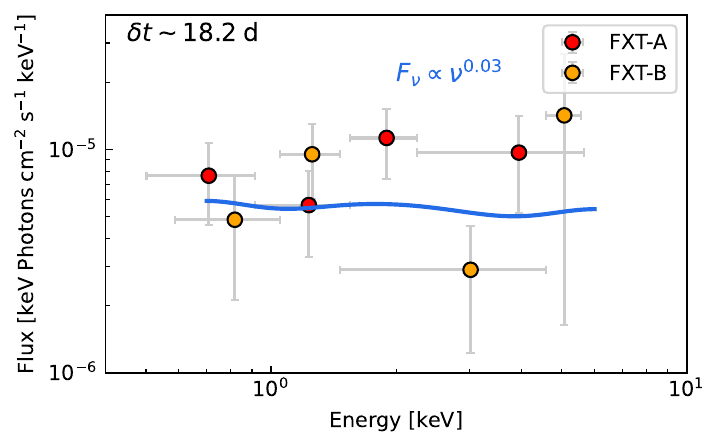}
        \put(-2,61){\textbf{b}}
        \end{overpic}
\end{minipage}
\begin{minipage}[b]{0.325\textwidth}
    \centering
        \begin{overpic}[width=1\textwidth]{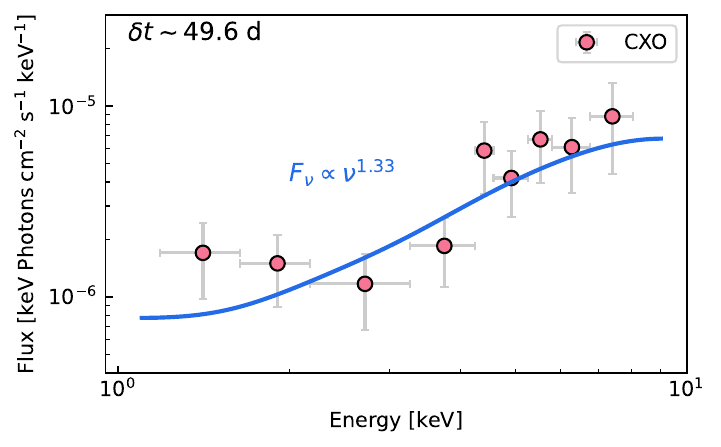}
        \put(-2,61){\textbf{c}} 
        \end{overpic}
\end{minipage}
\caption{
\noindent\textbf{The X-ray spectral evolution of AT 2024wpp in the 0.5--10 keV band.} 
		The panels display spectra at three representative epochs from the observations of FXT-A (red dots), FXT-B (orange dots), and {\it CXO} (pink dots).
		The blue lines represent the best fits of a power-law model with the Galactic neutral hydrogen absorption.
		The spectra show a clear hardening trend over time: 
		(a) At $\delta t \approx 5.2$ days, the {\it EP}-FXT observation yields a spectrum of $F_{\nu} \propto \nu^{-0.97}$.
		(b) At $\delta t \approx 18.2$ days, the {\it EP}-FXT data show a nearly flat spectrum, described by $F_{\nu} \propto \nu^{-0.03}$.
		(c) At $\delta t \approx 49.6$ days, the {\it CXO} data reveal an inverted spectrum with $F_{\nu} \propto \nu^{1.33}$.
}
\label{fig:XraySpectrum}
\end{figure}

\begin{figure}
	\centering
	\includegraphics[width=0.8\columnwidth]{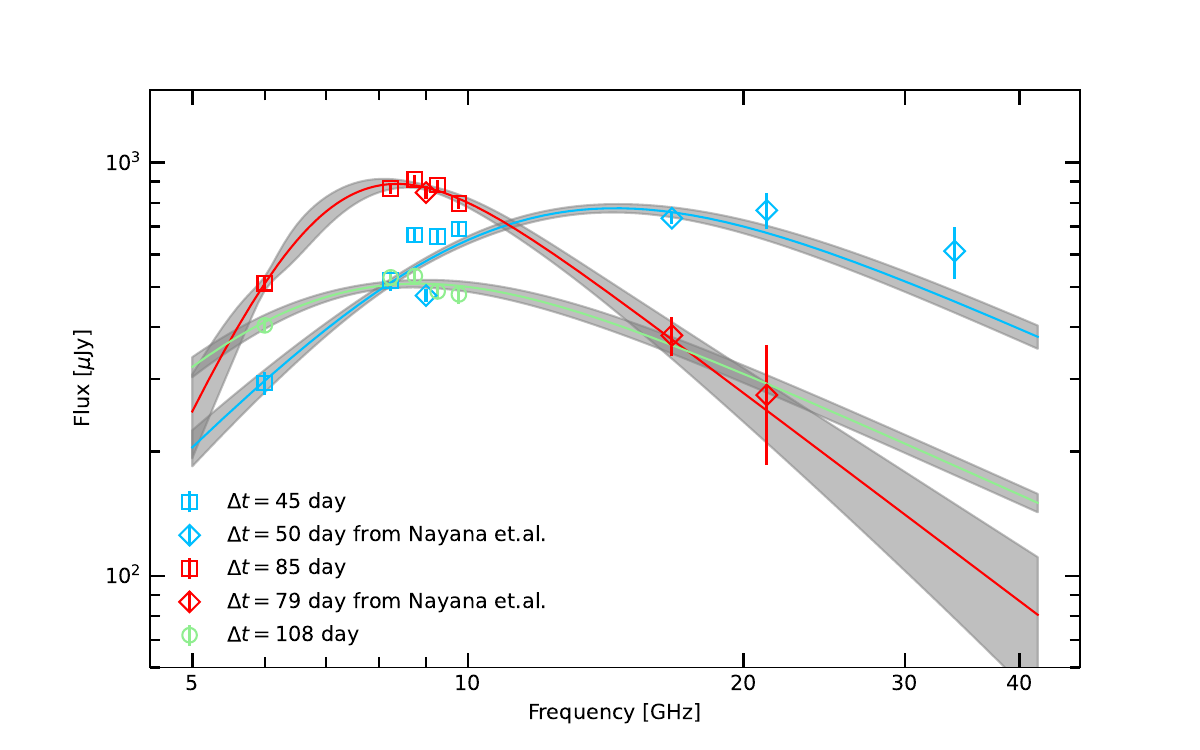}
    \caption{The radio SED of AT 2024wpp. The solid lines indicate the best-fit broken power-law model, and the grey shaded areas denote the corresponding confidence intervals at $1\sigma$.}
    \label{fig:radio_sed}
\end{figure}

\begin{figure}
	\centering
	\includegraphics[width=0.8\columnwidth]{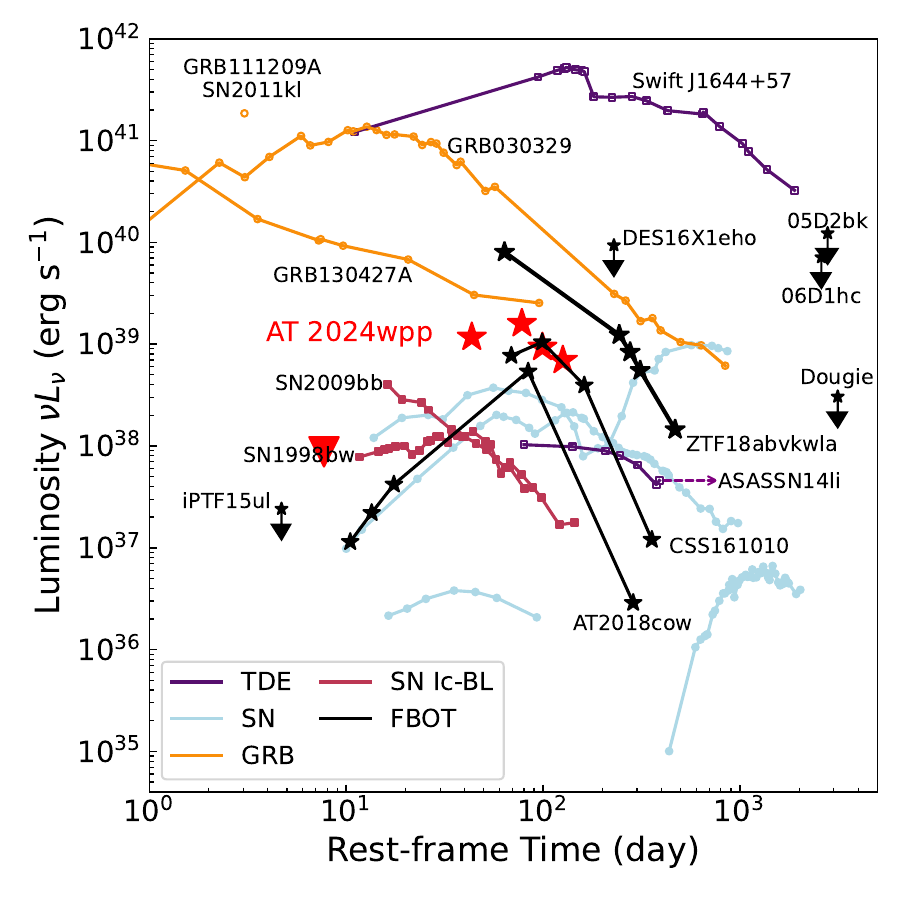}
    \caption{The 9 GHz Radio light curve of AT 2024wpp. The red triangle and stars denote the upper limit and the detected luminosities of AT 2024wpp, respectively. The low-frequency (1--10 GHz) light curves of different classes of energetic explosions: TDEs, SNe, relativistic SNe~Ic-BL, long-duration GRBs, and FBOTs
    are plotted for comparison.}
    \label{fig:radio}
\end{figure}

\begin{figure}
    \centering
	\includegraphics[width=0.8\columnwidth]{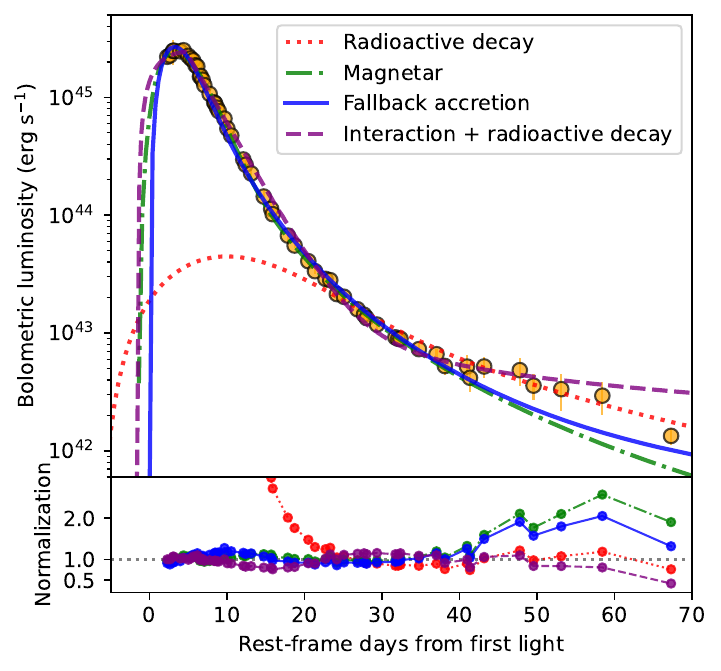}
    \caption{Model fitting results for the bolometric light curves of AT 2024wpp (yellow circles). Upper panel shows theoretical light curves from the radioactive decay model (red-dotted line), magnetar model (green dashed-dotted line), fallback accretion model (blue line), and interaction + radioactive decay model (purple dashed line). Lower panel shows the observed light curves normalized by above models, respectively.}
    \label{fig: model_sim}
\end{figure}

\begin{figure}
    \centering
	\includegraphics[width=0.8\columnwidth]{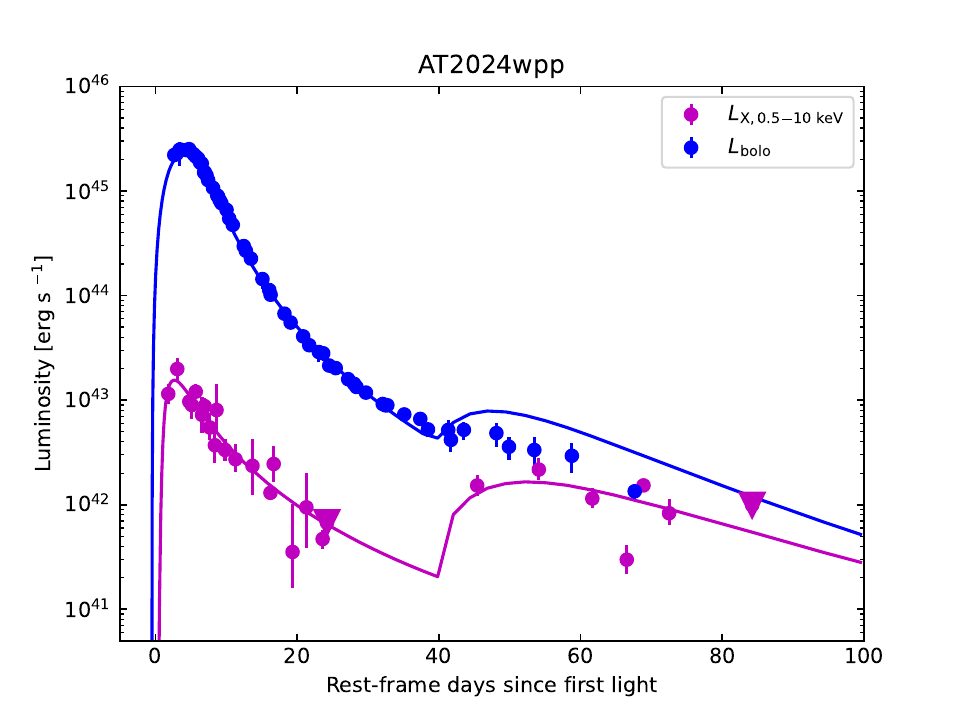}
    \caption{The accreting magnetar model fitted to the luminosity evolution of AT 2024wpp, including the bolometric luminosity (blue) and the soft X-ray luminosity (0.5--10 keV; purple). Filled circles show detections, while the inverted triangles indicate upper limits in the 0.5--10 keV band. Solid lines represent the corresponding model curves.}
    \label{fig:models_m}
\end{figure}

\begin{figure}[htbp]
\centering
\begin{minipage}[c]{1\textwidth}
    \centering
        \begin{overpic}[width=0.6\textwidth]{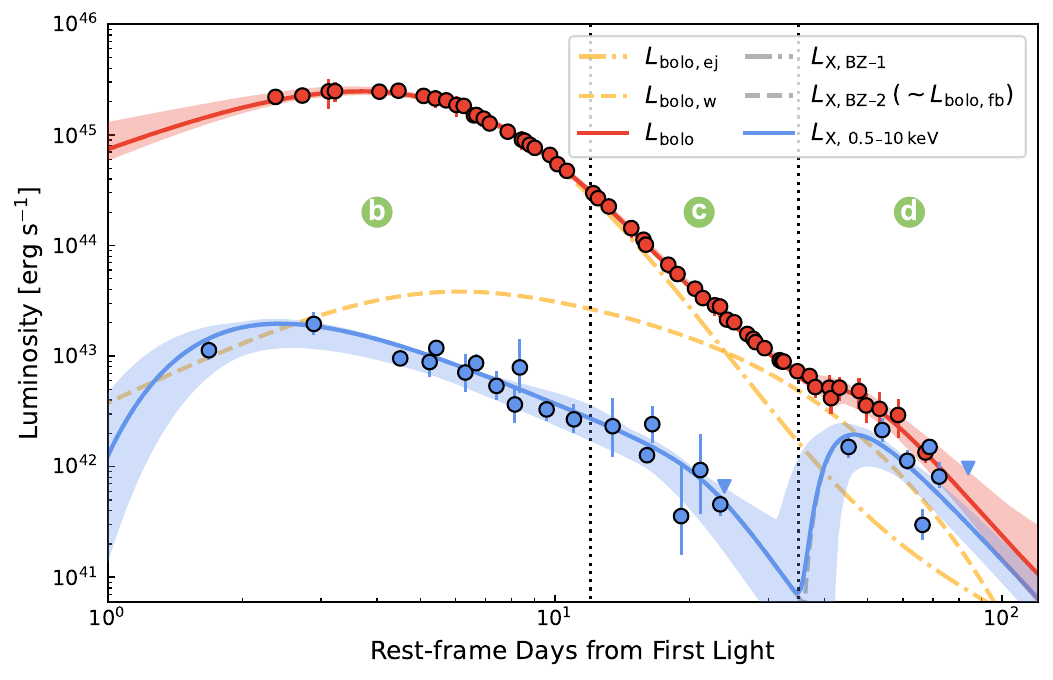}
        \put(0,62){\textbf{a}} 
        \end{overpic}
\end{minipage}
\begin{minipage}[c]{0.52\textwidth}
        \begin{overpic}[width=1\textwidth]{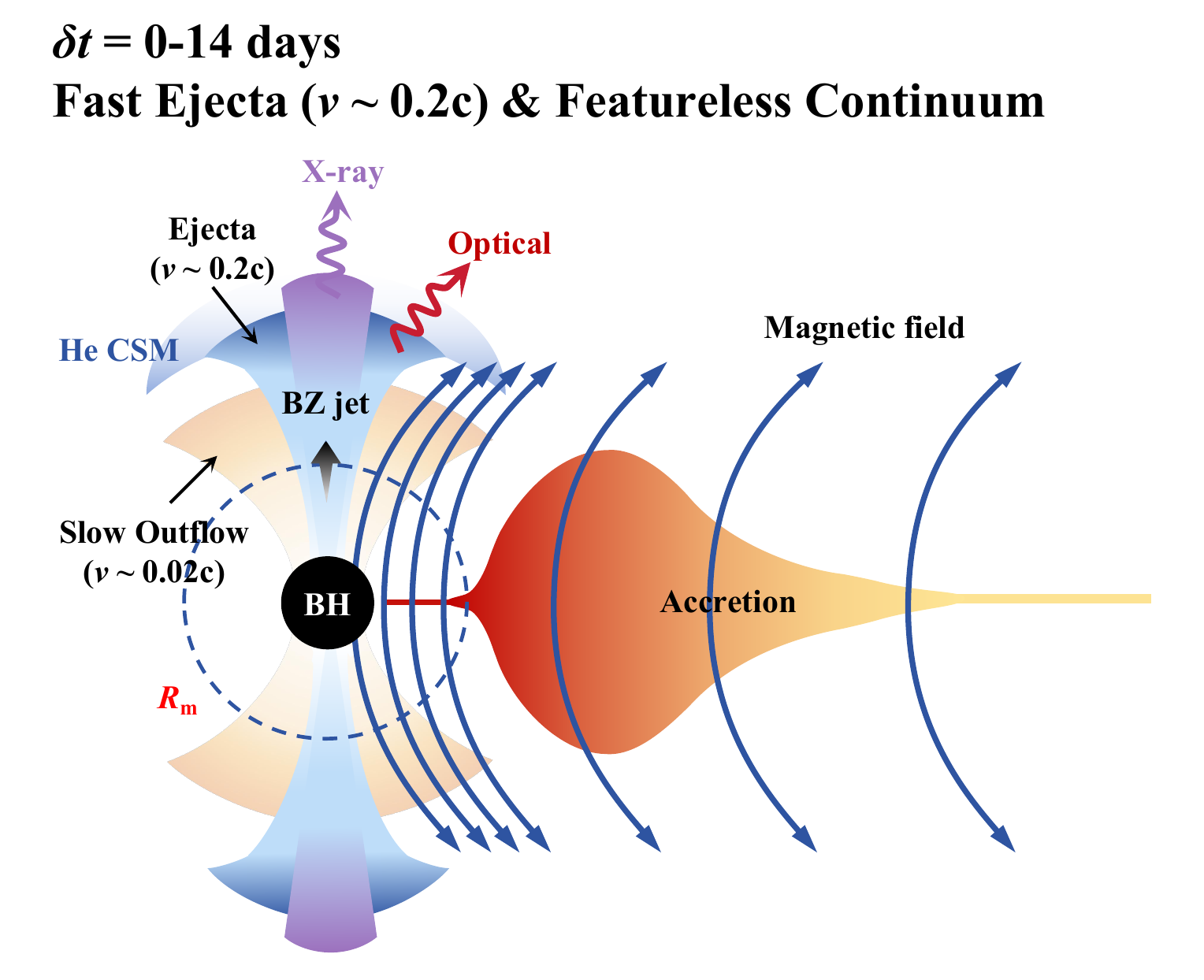}
        \put(-1,76){\textbf{b}}
        \end{overpic}
\end{minipage}
\begin{minipage}[c]{0.32\textwidth}
    \centering
        \begin{overpic}[width=1\textwidth]{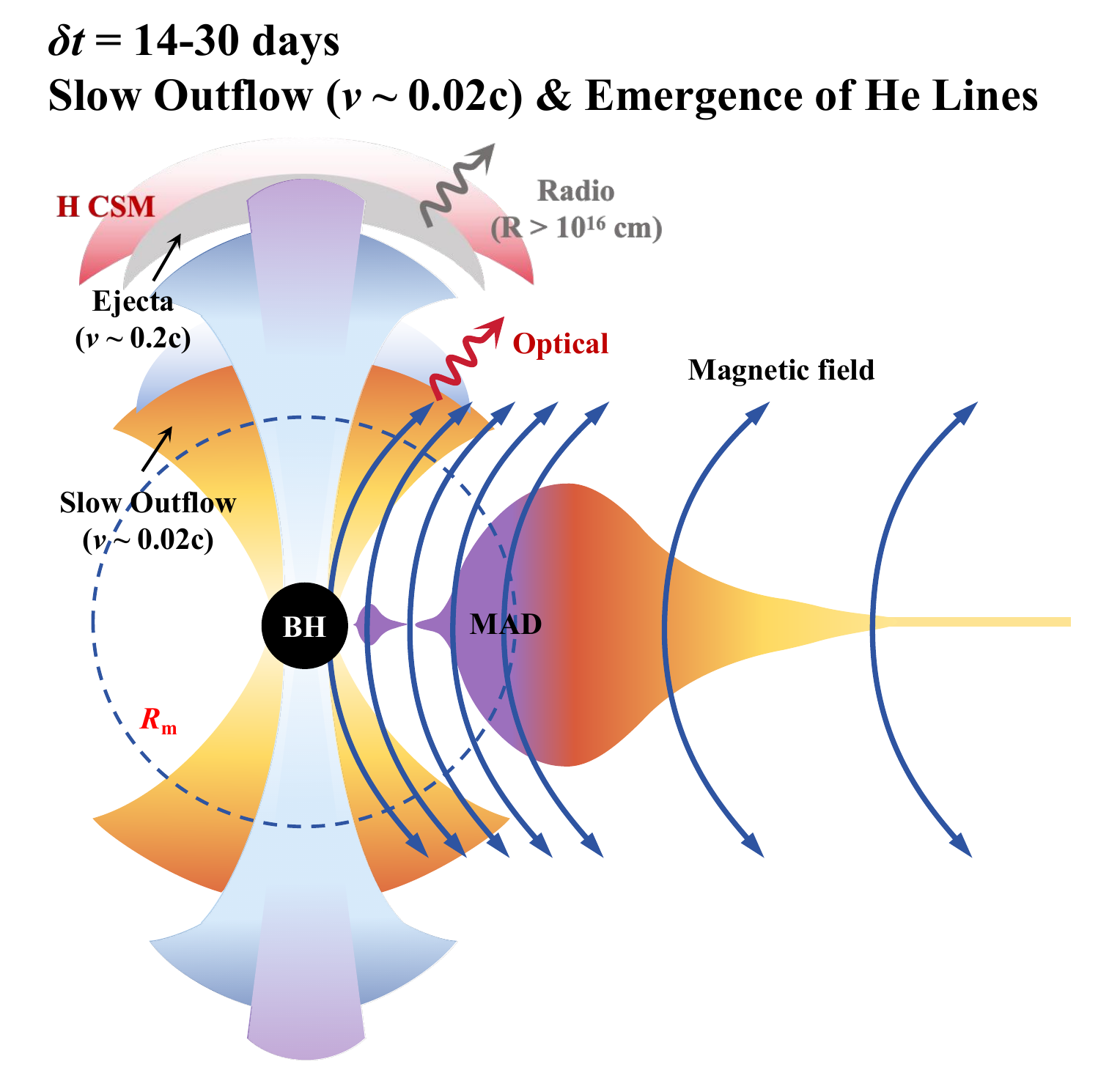}
        \put(-2,90){\textbf{c}} 
        \end{overpic}
        \begin{overpic}[width=1\textwidth]{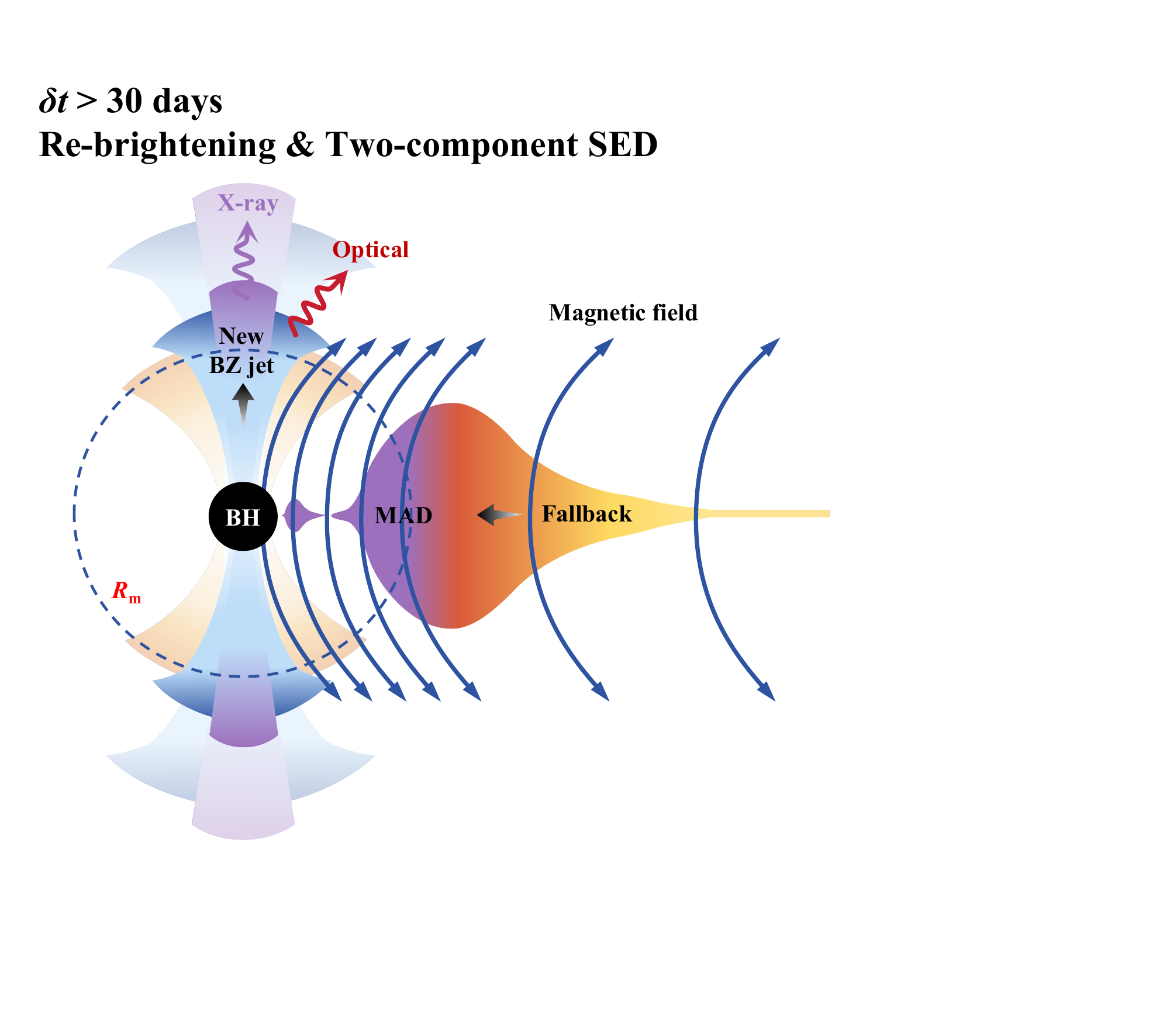}
        \put(-2,80){\textbf{d}} 
        \end{overpic}
\end{minipage}
\caption{
\noindent\textbf{Model-fit results to the bolometric and X-ray luminosity evolution of AT 2024wpp.} 
}
\label{fig:Model}
\end{figure}

\clearpage
\addcontentsline{lof}{figure}{\protect\numberline{\thefigure}\textbf{Model-fit results to the bolometric and X-ray luminosity evolution of AT 2024wpp.}}
\begingroup
\centering
\textbf{Figure \thefigure: \textbf{Model-fit results to the bolometric and X-ray luminosity evolution of AT 2024wpp.}
        (a) The model fit of a WR star and a BH merger to the bolometric (red line) and X-ray (blue line) light curves of AT 2024wpp, same as Fig.\,\ref{fig:model fit}(b). The three evolutionary phases corresponding to panels (b)--(d) are marked by green circles and separated by vertical dotted lines. 
		(b) Fast ejecta and jet-powered phase: after the WR-BH merger, the spinning BH generates a persistent jet powered by the BZ mechanism, producing the initial luminous optical and X-ray emission. 
        The optical emission originates from the X-ray reprocessing and shock interaction with the He-rich CSM. The outflow is anisotropic, with a high-velocity component in the polar regions and a slow component near the equator.
		(c) Slow outflow onset and jet-fading phase: the slow outflow becomes radiatively visible, as indicated by the emergence of narrow He lines with a blueshift of $v\approx 0.02c$. 
        As the accretion rate declines, the inner disk transitions into a MAD state, and the magnetic pressure pushes the magnetosphere outward, reducing the intensity of magnetic field and causing the X-ray emission to fade rapidly.
        Meanwhile, the ejecta spread to $R \approx 10^{16}\,\mathrm{cm}$, where they interact with the extended H-rich CSM and produce the radio synchrotron emission.
		(d) Fallback and rebrightening phase: gravitationally bound pre-explosion equatorial outflow falls back toward the BH, forming a new BZ jet. This fallback accretion powers the observed X-ray rebrightening and the slight optical bump.}
\endgroup
\clearpage


\begin{table}
\centering
\caption{Photometry of AT~2024wpp}\label{table:all_lc}
\begin{tabular}{ccccc}
\hline
\hline
MJD & Magnitude & Error$^a$ & Filter & Source \\
day &     mag   & mag   &        &    \\
\hline
60578.436 & $>20.085$ & & $g$ & ZTF \\
60579.059 & 18.320 & 0.080 & $L$ & GOTO \\
60579.226 & 17.423 & 0.185 & $g$ & ASASSN \\
60579.371 & 17.558 & 0.052 & $g$ & ZTF \\
60579.431 & 17.959 & 0.060 & $r$ & ZTF \\
60579.556 & 17.660 & 0.010 & $w$ & P1 \\
60579.557 & 17.207 & 0.191 & $g$ & ASASSN \\
60580.002 & 17.603 & 0.041 & $o$ & ATLAS \\
60580.005 & 17.604 & 0.043 & $o$ & ATLAS \\
... & ... & ... & ... & ... \\
60651.943 & 22.690 & 0.095 & $g$ & GTC \\
60651.949 & 22.755 & 0.113 & $r$ & GTC \\
60651.952 & 22.729 & 0.158 & $i$ & GTC \\
\hline
\end{tabular}
\begin{flushleft}
\centering
{$^a$ Uncertainty of magnitude, 1 $\sigma$.}
\end{flushleft}
\end{table}

\begin{table}
\centering
\caption{Overview of optical spectra of AT~2024wpp}\label{table:spec}
\begin{tabular}{cccccc}
\hline
\hline
MJD & Date & Phase$^a$ & Range & Exp. & Instrument/Telescope \\
(day) &      & (day)       & (\AA)        & (s)    & \\
\hline
60585.7 & 20241002 & 6.4 & 3770-8912 & 3000.0 & BFOSC/XLT \\
60588.9 & 20241005 & 9.3 & 3583-7456 & 2400.136 & RSS/SALT \\
60590.0 & 20241007 & 10.4 & 3802-7231 & 1800.0 & AFOSC/Copernico \\
60590.1 & 20241007 & 10.5 & 4005-9167 & 1200.0 & OSIRIS/GTC \\
60594.1 & 20241011 & 14.2 & 4007-9008 & 400.0 & OSIRIS/GTC \\
60595.8 & 20241012 & 15.6 & 3614-8887 & 3299.974 & YFOSC/LJT \\
60612.1 & 20241029 & 30.7 & 4010-9203 & 1800.0 & OSIRIS/GTC \\
60620.0 & 20241106 & 38.0 & 3993-9167 & 900.0 & OSIRIS/GTC \\
60638.4 & 20241124 & 54.9 & 3479-10352 & 2894.31 & LRISpBLUE/Keck I \\
60649.4 & 20241205 & 65.0 & 3061-10273 & 2708.86 & LRIS/Keck I \\
\hline
\end{tabular}
\begin{flushleft}
\centering
{$^a$ Relative to the time of first light, MJD 60578.750.}
\end{flushleft}
\end{table}

\begin{table}
\centering
\caption{Blackbody fitting results of AT~2024wpp}\label{table:blackbody}
\begin{tabular}{llllll}
\hline
\hline
MJD & Temperature & Luminosity & MJD & Temperature & Luminosity \\
day & ($10^3$ K) & ($10^{43}$ erg s$^{-1}$) & (day) & ($10^3$ K) & ($10^{43}$ erg s$^{-1}$) \\
\hline
60581.328 & $37.1_{-1.3}^{+1.3}$ & $221_{-18}^{+18}$ & 60596.102 & $24.2_{-1.4}^{+1.4}$ & $10.2_{-1.2}^{+1.3}$ \\
60581.710 & $36.6_{-1.4}^{+1.4}$ & $227_{-19}^{+23}$ & 60598.226 & $23.0_{-1.9}^{+1.9}$ & $6.7_{-1.2}^{+1.2}$ \\
60582.140 & $37.2_{-3.7}^{+3.7}$ & $248_{-59}^{+90}$ & 60599.173 & $22.40_{-0.84}^{+0.84}$ & $5.53_{-0.40}^{+0.30}$ \\
60582.252 & $37.2_{-2.9}^{+2.9}$ & $249_{-43}^{+56}$ & 60601.098 & $21.9_{-1.0}^{+1.0}$ & $4.07_{-0.33}^{+0.30}$ \\
60583.151 & $36.9_{-1.1}^{+1.1}$ & $246_{-16}^{+19}$ & 60602.024 & $21.12_{-0.79}^{+0.79}$ & $3.35_{-0.20}^{+0.24}$ \\
60583.605 & $37.5_{-1.9}^{+1.9}$ & $251_{-28}^{+26}$ & 60603.502 & $21.7_{-1.9}^{+1.9}$ & $2.88_{-0.46}^{+0.66}$ \\
60584.276 & $36.7_{-1.3}^{+1.3}$ & $226_{-19}^{+24}$ & 60604.170 & $22.4_{-2.7}^{+2.7}$ & $2.81_{-0.54}^{+0.56}$ \\
60584.619 & $36.4_{-2.0}^{+2.0}$ & $214_{-28}^{+50}$ & 60605.092 & $20.21_{-0.71}^{+0.71}$ & $2.14_{-0.13}^{+0.24}$ \\
60584.959 & $36.6_{-1.7}^{+1.7}$ & $206_{-22}^{+25}$ & 60606.090 & $20.8_{-1.5}^{+1.5}$ & $2.03_{-0.25}^{+0.22}$ \\
60585.293 & $35.7_{-2.3}^{+2.3}$ & $187_{-31}^{+50}$ & 60607.997 & $20.3_{-1.8}^{+1.8}$ & $1.59_{-0.21}^{+0.28}$ \\
60585.543 & $36.0_{-1.6}^{+1.6}$ & $183_{-19}^{+23}$ & 60608.904 & $19.8_{-1.2}^{+1.2}$ & $1.43_{-0.14}^{+0.15}$ \\
60585.906 & $34.0_{-1.8}^{+1.8}$ & $151_{-16}^{+23}$ & 60609.265 & $19.24_{-0.78}^{+0.78}$ & $1.338_{-0.095}^{+0.13}$ \\
60586.002 & $34.4_{-1.6}^{+1.6}$ & $152_{-14}^{+20}$ & 60610.720 & $18.9_{-1.2}^{+1.2}$ & $1.18_{-0.14}^{+0.15}$ \\
60586.282 & $33.82_{-0.68}^{+0.68}$ & $140.6_{-7.0}^{+7.6}$ & 60613.301 & $18.27_{-0.99}^{+0.99}$ & $0.919_{-0.068}^{+0.075}$ \\
60586.514 & $32.9_{-1.2}^{+1.2}$ & $127_{-11}^{+17}$ & 60613.701 & $18.1_{-1.1}^{+1.1}$ & $0.893_{-0.090}^{+0.083}$ \\
60587.270 & $32.5_{-1.2}^{+1.2}$ & $107.1_{-9.2}^{+10}$ & 60614.006 & $18.55_{-0.99}^{+0.99}$ & $0.894_{-0.062}^{+0.049}$ \\
60587.913 & $32.3_{-2.6}^{+2.6}$ & $91_{-13}^{+22}$ & 60616.601 & $17.8_{-1.4}^{+1.4}$ & $0.730_{-0.093}^{+0.12}$ \\
60588.043 & $32.1_{-1.6}^{+1.6}$ & $89_{-10}^{+8.7}$ & 60619.059 & $19.0_{-1.7}^{+1.7}$ & $0.661_{-0.084}^{+0.078}$ \\
60588.293 & $31.6_{-1.3}^{+1.3}$ & $81.7_{-8.0}^{+7.8}$ & 60620.237 & $19.2_{-1.5}^{+1.8}$ & $0.53_{-0.10}^{+0.10}$ \\
60588.543 & $31.8_{-2.1}^{+2.1}$ & $77_{-10}^{+9.5}$ & 60623.346 & $19.7_{-1.7}^{+2.1}$ & $0.52_{-0.16}^{+0.16}$ \\
60589.342 & $32.1_{-2.2}^{+2.2}$ & $66.0_{-9.5}^{+8.0}$ & 60623.757 & $19.8_{-1.7}^{+2.1}$ & $0.42_{-0.12}^{+0.12}$ \\
60589.756 & $30.9_{-1.0}^{+1.0}$ & $54.4_{-4.0}^{+4.1}$ & 60625.744 & $20.2_{-1.9}^{+2.3}$ & $0.52_{-0.13}^{+0.13}$ \\
60590.303 & $30.5_{-1.7}^{+1.7}$ & $47.4_{-6.0}^{+8.1}$ & 60630.735 & $21.0_{-2.2}^{+2.7}$ & $0.48_{-0.15}^{+0.15}$ \\
60591.984 & $28.9_{-1.6}^{+1.6}$ & $29.7_{-3.2}^{+3.3}$ & 60632.661 & $21.4_{-2.3}^{+2.9}$ & $0.36_{-0.11}^{+0.11}$ \\
60592.308 & $28.0_{-1.6}^{+1.6}$ & $26.8_{-2.9}^{+3.2}$ & 60636.546 & $22.1_{-2.6}^{+3.2}$ & $0.33_{-0.14}^{+0.14}$ \\
60593.085 & $27.7_{-1.6}^{+1.6}$ & $22.6_{-3.0}^{+3.6}$ & 60642.309 & $21.0_{-2.3}^{+2.9}$ & $0.29_{-0.11}^{+0.11}$ \\
60594.847 & $26.2_{-2.6}^{+2.6}$ & $14.4_{-2.1}^{+3.3}$ & 60651.943 & $18.4_{-2.0}^{+2.5}$ & $0.135_{-0.027}^{+0.027}$ \\
60595.892 & $25.2_{-2.0}^{+2.0}$ & $11.3_{-1.4}^{+2.0}$ &  & &  \\
\hline
\end{tabular}
\end{table}

\begin{table}
\centering
\caption{X-ray Observations of AT~2024wpp with {\it EP}-FXT, {\it Swift}-XRT, and {\it CXO}.}\label{table:specx}
\begin{tabular}{cccccc}
\hline
\hline
Start Date & Mid MJD & Exposure & Instrument & Flux (68\% CL) & $\Gamma$ (68\% CL) \\
(yy-mm-dd) &      &  (ks)   &    & ($10^{-13}$ $\mathrm{erg \,s^{-1} \,cm^{-2}}$)    & \\
\hline
			2024-09-27       & 60580.58  &  4.04 & Swift-XRT      & $5.69_{- 1.09} ^{+1.37} $         & $1.81_{-0.24} ^{+0.25} $           \\
			2024-09-28       & 60581.89  & 2.22 & Swift-XRT      & $9.84_{-2.04} ^{+2.65}$         & $1.88_{-0.29} ^{+0.30} $           \\
			2024-09-30       & 60583.65  &  3.09  & EP-FXT         & $4.80_{-0.60} ^{+0.73}$         & $1.97_{-0.17} ^{+0.18}$            \\
			2024-10-01       & 60584.45  & 5.84  & Swift-XRT      & $4.45_{-1.16} ^{+1.32}$            & $1.66_{-0.30} ^{+0.31}$           \\
			2024-10-01       & 60584.65  &  3.07 & EP-FXT         & $5.97_{-0.84}^{+1.03 }$           & $1.74_{-0.17} ^{+0.18}$           \\
			2024-10-02       & 60585.60  &  1.61  & Swift-XRT    & $3.58_{-1.18} ^{+1.71 }$            & $2.20_{-0.49} ^{+0.54}$        \\
			2024-10-02       & 60585.99  &  3.06  & EP-FXT         & $4.35_{-0.71} ^{+0.89}$           & $1.66_{-0.19} ^{+0.20}$           \\
			2024-10-03       & 60586.79  &  3.06 & EP-FXT         & $2.71_{-0.68} ^{+1.00}$           & $1.53_{-0.30} ^{+0.31}$           \\
			2024-10-04       & 60587.59  &  3.05  & EP-FXT         & $1.84_{-0.59} ^{+0.89}$          & $1.67_{-0.44} ^{+0.50}$         \\
			2024-10-04       & 60587.82  &  1.60  & Swift-XRT     & $3.98_{-1.64} ^{+3.16}$          & $2.02_{-0.68} ^{+0.72}$         \\
			2024-10-06       & 60589.16  &  6.09  & EP-FXT         & $1.66_{-0.35} ^{+0.46}$            & 
			$1.56 \pm 0.25$          \\
			2024-10-07       & 60590.73  &  3.04  & EP-FXT         & $1.35_{-0.33} ^{+0.52}$           & $2.01_{-0.40} ^{+0.42}$         \\
			2024-10-10       & 60593.37  &  4.10  & EP-FXT         & $1.17_{-0.55} ^{+0.94}$            & $0.91_{-0.49} ^{+0.55}$           \\
			2024-10-13       & 60596.21  &  19.81  & CXO        & $0.64_{-0.10} ^{+0.12}$          & $1.40_{-0.29} ^{+0.28}$            \\
			2024-10-13       & 60596.68  &  9.05  & EP-FXT         & $1.22_{-0.40} ^{+0.58}$           & $0.97_{-0.32} ^{+0.33}$          \\
			2024-10-16       & 60599.55 &  8.82  & EP-FXT         & $0.18_{-0.10} ^{+0.33}$           & $1.76_{-1.01} ^{+1.13}$           \\
			2024-10-18       & 60601.72  &  12.07  & EP-FXT         & $0.47_{-0.28} ^{+0.52}$           & $0.54_{-0.77} ^{+0.76}$          \\
			2024-10-21       & 60604.21  &  19.81  & CXO        & $0.23\pm 0.05$            				& $1.34$ $^{a}$                        \\
			2024-10-21       & 60604.89   &  13.53  & EP-FXT         & $<0.33$ $(3\sigma)$                  &  
			$-$              \\
			2024-11-13       & 60628.01  & 36.09 & CXO        & $0.76_{-0.16} ^{+0.20}$           & $-0.33_{-0.35} ^{+0.33}$                    \\
			2024-11-23       & 60637.33  &  6.10  & EP-FXT         & $1.08_{-0.23} ^{+0.31}$            & $1.81_{-0.28} ^{+0.29}$           \\
			2024-12-01       & 60645.39  &  10.12  & EP-FXT         & $0.57_{-0.10} ^{+0.15}$           & 
			$2.22\pm 0.34$                \\
			2024-12-06       & 60650.83   &  8.97 & EP-FXT         & $0.15_{-0.04} ^{+0.06}$         & $2.97_{-0.68} ^{+0.74}$            \\
			2024-12-09       &  60653.51  &  36.24  & CXO        & $0.76\pm 0.08$                       & 
			$1.60 \pm 0.25$                \\
			2024-12-13       & 60657.30  &  6.59  & EP-FXT         & $0.41_{-0.09} ^{+0.15}$          & $2.44_{-0.49} ^{+0.53}$             \\ 
			2024-12-25       & 60670.05   &  5.59  & EP-FXT         &  $<0.49$ $(3\sigma)$            &    
			$-$           \\
\hline
\end{tabular}
\begin{flushleft}
$^{a}$Owing to low photon counts, this photon index is fixed to the value obtained from the \texttt{tbabs*cflux*pow} model to ensure the fitting goodness.
\end{flushleft}
\end{table}

\begin{table}
\setlength{\tabcolsep}{10pt}
\small
    \centering
	\caption{Log of radio observations. }
	\label{tab2}
	\begin{tabular}{llllll} 
		\hline
        \hline
        Observed Date & Centroid MJD &Phase & Frequency & Flux Density & Exposure \\
        (yy.mm.dd) &  & (day) & (GHz) & ($\mu$Jy) & (hr) \\
		\hline

   24.10.04 & 60587.83  & 8.46 & 5.5  & $<41$ & 4 \\
   24.10.04 & 60587.83  & 8.46 & 9.0  & $<50$ & 4 \\
  24.11.12 & 60626.60 & 47.23 & 5.0 & $103\pm 16$ & 3\\
  24.11.12 & 60626.60 & 47.23 & 6.0 & $130\pm 12$ & 3\\
  24.11.12 & 60626.60 & 47.23 & 8.2 & $551\pm 10$ & 3\\
  24.11.12 & 60626.60 & 47.23 & 8.8 & $734\pm 31$ & 3\\
  24.11.12 & 60626.60 & 47.23 & 9.3 & $806\pm 29$ & 3\\
  24.11.12 & 60626.60 & 47.23 & 9.8 & $759\pm 30$ & 3\\
 24.12.20 & 60664.29 & 84.92 & 5.0 & $260\pm 12$ & 4\\
 24.12.20 & 60664.29 & 84.92 & 6.0 & $330\pm 6$ & 4\\
 24.12.20 & 60664.29 & 84.92 & 8.2 & $930\pm 19$ & 4\\
 24.12.20 & 60664.29 & 84.92 & 8.8 & $989\pm 18$ & 4\\
 24.12.20 & 60664.29 & 84.92 & 9.3 & $936\pm 16$ & 4\\
 24.12.20 & 60664.29 & 84.92 & 9.8 & $863\pm 20$ & 4\\
  25.01.12 & 60687.29 & 107.92 & 5.0 & $227\pm5$ & 4 \\
  25.01.12 & 60687.29 & 107.92 & 6.0 & $253\pm5$ & 4 \\
  25.01.12 & 60687.29 & 107.92 & 8.2 & $526\pm11$ & 4 \\
  25.01.12 & 60687.29 & 107.92 & 8.8 & $552\pm9$ & 4 \\
  25.01.12 & 60687.29 & 107.92 & 9.3 & $541\pm11$ & 4 \\
  25.01.12 & 60687.29 & 107.92 & 9.8 & $499\pm12$ & 4 \\
  25.02.10 & 60716.43 & 137.06 & 5.0 & $349\pm 79$ & 3\\
  25.02.10 & 60716.43 & 137.06 & 6.0 & $252\pm 68$ & 3\\
  25.02.10 & 60716.43 & 137.06  & 8.2 & $426\pm 79$ & 3\\
  25.02.10 & 60716.43 & 137.06 & 8.8 & $396\pm 40$ & 3\\
  25.02.10 & 60716.43 & 137.06 & 9.3 & $367\pm 54$ & 3\\
  25.02.10 & 60716.43 & 137.06  & 9.8 & $393\pm 50$ & 3\\

  \hline
  \hline
	\end{tabular}
\end{table}

\begin{table}
\setlength{\tabcolsep}{30pt}
\small
    \centering
	\caption{Parameters of broken power-law fits. }
	\label{tab1}
	\begin{tabular}{lccc} 
		\hline
        \hline
        Parameter &  $\Delta t=47$ day & $\Delta t=85$ day  & $\Delta t=108$ day\\
		\hline

  $F_{\nu, \text{peak}}$ ~($\mu$Jy) & $1431^{+31}_{-35}$ & $1564^{+99}_{-118}$ & $927^{+20}_{-19}$ \\
  $\nu_{\text{peak}}$  (GHz) & $11.2^{+0.6}_{-0.5}$ & $7.1^{+0.6}_{-0.4}$ & $6.8^{+0.3}_{-0.3}$ \\
  $\beta_{1}$ & $2.3^{+0.2}_{-0.2}$  & $5.0^{+1.6}_{-1.0}$ & $2.5^{\star}$ \\
 $\beta_{2}$ & $-1.0^{\star}$ & $-1.7^{+0.3}_{-0.3}$ & $-1.0^{\star}$ \\
 $R$~($10^{16}$ cm) & $4.4^{+0.2}_{-0.2}$ & $7.3^{+0.2}_{-0.3}$  & $6.0^{+0.2}_{-0.2}$\\
 $v$~($c$) & $0.39^{+0.01}_{-0.02}$ & $0.36^{+0.01}_{-0.02}$ &  $0.23^{+0.01}_{-0.01}$\\
 $B$~(G) & $0.75^{+0.04}_{-0.03}$ & $0.47^{+0.03}_{-0.03}$ & $0.48^{+0.02}_{-0.02}$ \\
 $n$~($10^2~{\rm cm^{-3}}$) & $4.5^{+1.0}_{-0.6}$  & $2.1^{+0.6}_{-0.3}$ & $1.0^{+0.2}_{-0.1}$ \\
 $U$~($10^{48}~{\rm erg}$) & $12.7^{+0.4}_{-0.5}$  & $22.1^{+0.7}_{-0.8}$ & $12.4^{+0.5}_{-0.4}$\\
 $\dot{M}$~($10^{-5}~{\rm M}_{\odot}~{\rm yr}^{-1}$) & $2.9^{+0.3}_{-0.2}$  & $1.1^{+0.2}_{-0.1}$ & $1.17^{+0.09}_{-0.08}$\\

 
  
  \hline
  \hline
	\end{tabular}
\begin{flushleft}
$^{\star}$ designates the parameters that are held fixed.
\end{flushleft}
\end{table}

\begin{table}
\setlength{\tabcolsep}{5pt}
\small
	\caption{Model parameters for radioactive decay, magnetar, fallback accretion, and  interaction + radioactive decay models. Here $\kappa=0.1$ cm$^{2}$ g$^{-1}$ and $v_\mathrm{ej}=0.2c$.}
	\begin{tabular}{lllll} 
		\hline
        \hline
        Parameter & radioactive decay & magnetar wind &  fallback accretion &  interaction + decay \\
		\hline
        $t_\mathrm{exp}$ & $-6.51_{-0.78}^{+0.77}$& $-1.48_{-0.42}^{+0.36}$ &$-0.0129_{-0.0237}^{+0.0098}$  & $-1.96_{-1.08}^{+0.87}$\\
        M$_\mathrm{ej}$ (M$_\odot$)& $11.9_{-0.4}^{+0.4}$ & $1.7_{-0.17}^{+0.2}$ & $0.72_{-0.08}^{+0.09}$ & $1.92_{-0.14}^{+0.19}$\\
        $f_\mathrm{Ni}$ & $0.199_{-0.001}^{+0.001}$ & -- &-- $0.19_{-0.02}^{+0.01}$\\
        $\kappa_\gamma$ (cm$^2$ g$^{-1}$) & $0.027$(fixed) &$0.00295_{-0.00065}^{+0.00168}$ &-- & --\\
        $P_0$ (ms ) & --&$1.14_{-0.1}^{+0.19}$&--&--\\
        $B$ ($10^{14}$ G) & -- &$0.69_{-0.09}^{+0.19}$&--&-- \\
        $L_\mathrm{0,acc}$ ($10^{46}$ erg s$^{-1}$) & --& --& $5.95_{-2.18}^{+4.7}$&--\\
        $t_\mathrm{d,acc}$ (day)& --&--&$0.598_{-0.205}^{+0.253}$& -- \\
        $E_\mathrm{SN}$ ($10^{51}$ erg)&--&--&--&$4.03_{-0.39}^{+0.39}$\\
        $M_\mathrm{CSM}$ (M$_\odot$)& --&--&--& $0.794_{-0.102}^{+0.141}$\\
        $\rho_{\mathrm{CSM},0}$ ($10^{-11}$ g cm$^{-3}$)& --&--&--&$0.259_{-0.195}^{+2.12}$\\
        $r_{\mathrm{CSM},0}$ ($10^{14}$ cm) & --&--&--& $2.28_{-1.59}^{+3.21}$\\
        $x_0$ & --&--&--& $0.318_{-0.171}^{+0.322}$\\
 \hline
        $\chi_\mathrm{opt}^2/$dof &2314/52& 29/50 & 32/51 & 139/47\\
  \hline
	\end{tabular}
\label{tab: model_sim}
\end{table}

\begin{table}[htbp]
\centering
\caption{Free parameters, priors, and best-fitting results in the accreting magnetar model.}
\label{tab:pars}
\begin{tabular}{lcr}
\hline\hline
Parameter & Prior & Result \\
\hline
$M_{\rm ej}~({\rm M}_\odot)$	&	$[0.01,1]$	&	$0.66^{+0.06}_{-0.04}$	\\
$v_{\rm SN}~(10^9~\rm cm~s^{-1})$	&	$[0.1,10]$	&	$4.10^{+1.46}_{-1.73}$	\\
$\log_{10}[B_p~(\rm G)]$	&	$[10,16]$	&	$14.86^{+0.74}_{-0.77}$	\\
$\log_{10}[P_0~(\rm s)]$	&	$[-3,-1]$	&	$-2.34^{+0.13}_{-0.19}$	\\
$\epsilon_e$	&	$[0.01,0.5]$	&	$0.11^{+0.07}_{-0.06}$	\\
$\log_{10}(\Gamma_{\rm sat})$	&	$[1,5]$	&	$2.91^{+0.37}_{-0.93}$	\\
$p$	&	$(2,3)$	&	$2.73^{+0.11}_{-0.26}$	\\
$t_{\rm expl}~(\rm day)$	&	$[0,10]$	&	$0.50^{+0.35}_{-0.31}$	\\
$\log_{10}[\dot{M}_0~({\rm M}_\odot~\rm s^{-1})]$	&	$[-10,-2]$	&	$-5.85^{+0.13}_{-0.20}$	\\
$\log_{10}[{t_0}~(\rm s)]$	&	$[6,7]$	&	$6.55^{+0.13}_{-0.18}$	\\
\hline
$\chi^2_{\mathrm{tot}}/\mathrm{dof}$	&	--	&	$236.53/67$	\\
\hline
\end{tabular}

Note: The uncertainties of the best-fitting parameters are measured as $1\sigma$ confidence ranges.
\end{table}

\begin{table}
\setlength{\tabcolsep}{10pt}
	\centering
		\caption{The parameters, priors, and MCMC constraint results in the WR-BH delayed merger model.}
			\label{tab:fit_results}
		\begin{tabular}{lcr}
			\hline \hline
			Parameters & Priors & Results ($1\sigma$)  \\
			\hline
			$M_{\mathrm{star}}\,({\rm M}_\odot)$ & $[0, 50]$ & $34^{+10}_{-10}$ \\
			$M_{\mathrm{BH}} \,({\rm M}_\odot)$ & $[0, 30]$ & $15.4^{+5.2}_{-8.8}$ \\
			$M_{\mathrm{ej}} \,({\rm M}_\odot)$ & $[0, 1]$ & $0.271 ^{+0.075}_{-0.066}$ \\
			$v_{\mathrm{ej}} \,(c)$ & $[0, 1]$ & $0.249 ^{+0.041}_{-0.033}$ \\
            $\log_{10}[M_{\mathrm{w}}\, ({\rm M}_\odot)]$ & $[-2, 0]$ & $-1.28^{+0.46}_{-0.34}$ \\
            $v_{\mathrm{w}} \,(c)$ & $[0, 0.1]$ & $0.0320 ^{+0.0044}_{-0.0038}$ \\
			$\log_{10}[\dot{M}_\mathrm{esc}\,({\rm M}_{\odot}\,  \mathrm{yr^{-1}})]$ & $[-3, 2]$ & $-0.16^{+0.12}_{-0.14}$  \\
            $t_\mathrm{esc}\,(\mathrm{days})$ & $[10, 200]$ & $43 ^{+5}_{-6}$  \\
			$\log_{10}[\eta_{\mathrm{acc}}]$ &$ [-3, 0] $& $-1.77 ^{+0.84}_{-1.10}$ \\			
			$\log_{10}[\eta_{\mathrm{BZ}}]$ & $[-4, -1] $& $-2.45 ^{+0.24}_{-0.30}$  \\
            $\log_{10}[B_H\,(\mathrm{G})]$ & $[11.5, 12.5]$ & $11.628 ^{+0.055}_{-0.120}$  \\
			$k$ &$ [0, 2.2] $& $1.77^{+0.17}_{-0.11}$ \\
			$t_{\rm fb}\, (\mathrm{days})$ &$ [20, 40] $& $30^{+7}_{-5}$ \\
			$\log_{10}[\dot{M}_{\mathrm{fb}}\, ({\rm M}_\odot  \, \mathrm{yr^{-1}})]$ & $[-3, 2] $ & $-1.38^{+0.52}_{-0.52}$ \\
			\hline
			$\chi^2_{\mathrm{tot}}/\mathrm{dof}$ & -- & $84.45/65$ \\
			\hline        
		\end{tabular}
\end{table}


\clearpage 

\paragraph{Caption for Data S1.}
\textbf{UV/optical photometric observations of AT~2024wpp.}
First column: observational MJD with a unit of day; second column: observed magnitude with a unit of mag;
third column: uncertainty of magnitude with a unit of mag; fourth column: observed bandpass;
fifth column: source of the data.



\end{document}